\documentclass[floatfix,preprintnumbers,citeautoscript,newabstract,nofootinbib]{revtex4} 
\pdfoutput=1
\usepackage[utf8]{inputenc}
\usepackage{amsmath,amssymb,graphics}
\usepackage[DIV12]{typearea}
\usepackage{amsmath,amsfonts,amssymb}
\usepackage{graphicx}
\usepackage{units}
\usepackage{mathrsfs}
\usepackage{bbm}
\usepackage{pifont}
\usepackage{braket}
\usepackage{units}
\usepackage[small,loose]{subfigure}  
\usepackage[dvipsnames]{xcolor}
\usepackage{mathtools}
\usepackage{tikz}
\usepackage[bookmarks=false]{hyperref}
\usepackage{enumerate}
\usepackage[normalem]{ulem}
\usepackage{cancel}
\usepackage{graphicx}

\newcommand{\Eqref}[1]{Equation~\eqref{#1}}
\newcommand{\Figref}[1]{Figure~\ref{#1}}

\newcommand{\eVdist}{\kern-0.06em}

\newcommand{\ev}{\:\text{e\eVdist V}}   

\newcommand{\cm}{\:\text{cm}}

\newcommand{\TS}{\ensuremath{T_\mathrm{S}}}%
\newcommand{\TSEff}{\ensuremath{\overline{T}_\mathrm{S}}}%
\newcommand{\Li}[2]{\ensuremath{\mathrm{Li}_{#1}(#2)}}
\newcommand{\lth}{\ensuremath{\lambda_\mathrm{th}}}

\allowdisplaybreaks[1]

\def\mytitle{Massive Fermi Gas in the Expanding Universe} 

\title{\boldmath\mytitle\unboldmath}

\begin{document}

\preprint{TUM-HEP 1061/16\\}

\title{\mytitle}

\author{Andreas Trautner}
\email[]{atrautner@uni-bonn.de}
\affiliation{Bethe Center for Theoretical Physics and Physikalisches Institut der Universit\"at Bonn,
Nussallee 12, 53115 Bonn, Germany}
\affiliation{Excellence Cluster Universe, Technische Universit\"at M\"unchen, Boltzmannstra\ss e~2, 85748 Garching, Germany
}

\date{\today}

\begin{abstract}
The behavior of a decoupled ideal Fermi gas in a homogeneously expanding three--dimensional volume is investigated, starting from 
an equilibrium spectrum.
In case the gas is massless and/or completely degenerate, the spectrum of the gas can be described 
by an effective temperature and/or an effective chemical potential, both of which scale down with the volume expansion. 
In contrast, the spectrum of a decoupled massive and non--degenerate gas can only be described by an effective temperature
if there are strong enough self--interactions such as to maintain an equilibrium distribution.
Assuming perpetual equilibration, we study a decoupled gas which is relativistic at decoupling and then is red--shifted until it becomes non--relativistic.
We find expressions for the effective temperature and effective chemical potential which allow us 
to calculate the final spectrum for arbitrary initial conditions.
This calculation is enabled by a new expansion of the Fermi--Dirac integral, which is for our purpose superior to the well--known Sommerfeld expansion. 
We also compute the behavior of the phase space density under expansion and compare it to the case of real temperature 
and real chemical potential. Using our results for the degenerate case, we also obtain the mean relic velocity of the recently proposed non--thermal cosmic neutrino background.
\end{abstract}

\maketitle

\section{Introduction}

Let us consider a gas of particles which has been in thermal equilibrium with its surrounding until a given point (the ``freeze-out'' or ``decoupling''), and which is subsequently treated
as a gas of non--interacting, i.e.\ 
freely streaming particles in an expanding volume. 
Before decoupling we can assign a real temperature $T$ and chemical potential $\mu$ to the system, which parametrically enter the spectrum of the thermally coupled gas. 
After decoupling, the particles of the gas are non--interacting, implying that the notion of a thermal equilibrium and, hence, also the notion of a temperature is meaningless. 
Instead, the particle number is conserved.

It is a common lore that the form of the phase space distribution of particles (i.e.\ the form of the spectrum) in such a system is invariant under spatial expansion. 
That is, one may infer the ``red--shifted'' spectrum after decoupling by evaluating the original spectrum at an ``effective temperature'' which corresponds to the red--shifted, i.e.\ rescaled,
initial temperature. Depending on whether the gas is ultra--relativistic or non--relativistic \textit{at decoupling}, the effective temperature scales inversely proportional to the scale factor
or inversely proportional to its square, respectively (cf.\ e.g.\ \cite{Kolb:1990vq, Dodelson:2003ft, Mukhanov:2005sc, Weinberg:2008}).

The statement that the form of the spectrum is invariant under spatial expansion, however, is only exactly true in case 
the particles of the gas are massless; and it is true to a very good approximation as long as the particles of the gas are \textit{either} 
all highly relativistic \textit{or} all of them non--relativistic.
If, however, a massive gas decouples in a highly relativistic state and then is red--shifted until it becomes non--relativistic,
it depends on the details of interactions within the gas whether or not the spectrum changes.

Assuming a strictly non--interacting gas implies that the occupation number of each single mode cannot change.
This implies that the spectrum after expansion cannot be described by a common effective temperature which holds for all modes.\footnote{%
To be clear, the occupation number of each mode with momentum $p$ can in this case very well be approximated by the common expression $f(p,T)~=~[1+\mathrm{exp}(p/T)]^{-1}$ where
$T$ scales like the inverse of the scale factor; but $T$ here is \emph{not} an effective Fermi--Dirac temperature.\label{SMneutrinos}}
However, if there is self--interaction among the particles of the decoupled gas, then
a redistribution of particles among the different modes occurs while the total particle number is conserved.
If the self--interaction is sufficiently strong then an equilibrium Fermi--Dirac spectrum will be retained \cite{Bernstein:1988bw,Lesgourgues:2013}.

\begin{figure}[t]
\includegraphics[width=0.45\textwidth]{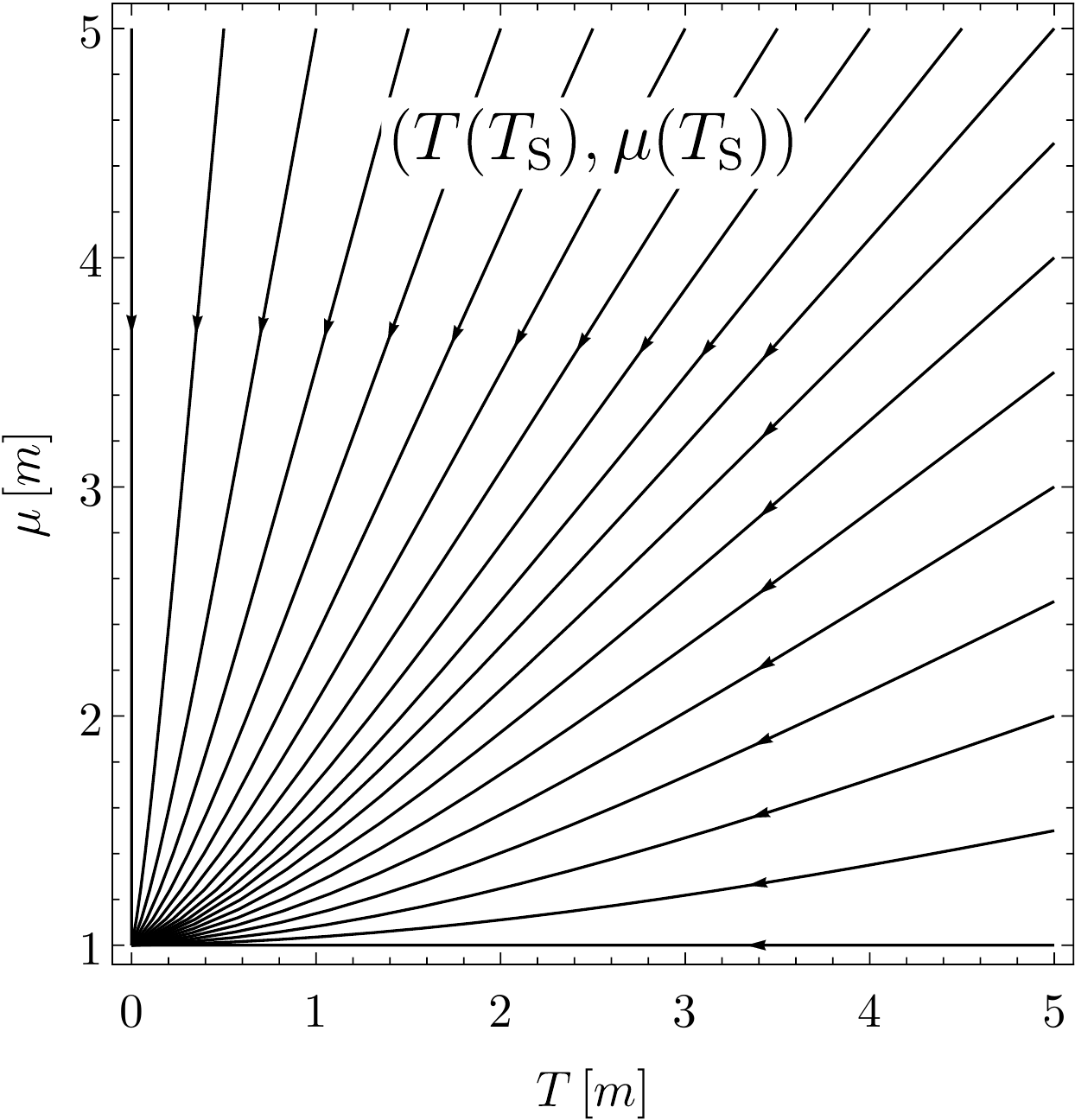}
\caption{Evolution of the effective temperature $T(\TS)$ and the effective chemical potential $\mu(\TS)$ of a decoupled massive ideal Fermi gas in dependence of the volume expansion
(here expressed in terms of the ``scaling temperature'' $\TS\propto 1/R$).
Indicated by the different lines is the evolution of $(T(\TS),\mu(\TS))$ in the expanding universe for several different initial conditions $(T_0,\mu_0)$.
Letting $\TS$ run from $T_0$ at decoupling to $\TS\sim T_\gamma$ today the gas evolves along the lines in the indicated direction. 
For the massless case, $T_0$ and $\mu_0$ are simply linearly rescaled and all lines would be straight.}
\label{fig:TMuScaling}
\end{figure}

In this work, we derive analytic formulae describing how the spectrum of a decoupled gas changes 
under volume expansion. This is done under the crucial assumption of (i) particle number conservation, and (ii) the preservation of 
an equilibrium Fermi--Dirac spectrum which depends only on the
two parameters $T_\mathrm{eff}=T(R)$ and $\mu_\mathrm{eff}=\mu(R)$, where $R$ is the scale factor. 
Our main result is \textit{how} the spectrum changes 
under volume expansion, i.e.\ the functions $T(R)$ and $\mu(R)$ (cf.\ \Figref{fig:TMuScaling}).

All results which we derive are general and could find applications for any decoupled but sufficiently strong self--interacting sector
such as, for example, massive self--interacting neutrinos\footnote{%
While non--standard neutrino self--interactions before and during the epoch of CMB formation are highly constrained \cite{Hannestad:2005ex, Basboll:2008fx, Archidiacono:2013dua}
it seems not to be observationally excluded that relic neutrinos recouple after CMB formation but before they turn non--relativistic.}, 
decoupled self--interacting Dark Matter \cite{Spergel:1999mh,Wandelt:2000ad,Harvey:2015hha,Bernal:2015ova,Chu:2016pew}, 
or self--interacting sterile neutrinos \cite{Hannestad:2013ana, Dasgupta:2013zpn, Bringmann:2013vra, Archidiacono:2014nda, Archidiacono:2016kkh}.

For the ease of the discussion, we will assume an instantaneous decoupling and not take into account distortions of the initial spectrum 
which could result from an ``incomplete decoupling'' (cf.\ e.g.\ \cite{Dolgov:2002wy, Mangano:2005cc}).
We treat cosmic expansion as strictly homogeneous and neglect metric perturbations for the time being (cf.\ e.g.\ \cite{Lesgourgues:2013,Ma:1995ey}).
We work in a large (compared to the Compton wavelength of the fermions) and isotropic three--dimensional volume 
and use a fully relativistic dispersion relation in order to obtain the density of states. The occupancy of each state is given by the familiar Fermi--Dirac distribution.
The number density $n$ of fermions per energy interval $\mathrm{d}E$ then is given by\footnote{%
We work in units $\hbar=c=k_\mathrm{B}=1$.}
\begin{equation}\label{eq:FermiSpectrum}
 \frac{\mathrm{d}n}{\mathrm{d}E}(T,\mu)~=~\frac{g}{2\pi^2}\frac{E\left(E^2-m^2\right)^{1/2}}{\mathrm{e}^{(E-\mu)/T}+1}\;,
\end{equation}
where $m$ is the fermion mass, $\mu$ the chemical potential, $T$ denotes the temperature, and $g$ counts internal degrees of freedom. The integrated spectrum 
\begin{equation}\label{eq:FermiDiracIntegral}
 n(T,\mu)~=~\frac{g}{2\pi^2}\,\int\displaylimits_m^\infty\mathrm{d}E\,\frac{E\left(E^2-m^2\right)^{1/2}}{\mathrm{e}^{(E-\mu)/T}+1}
\end{equation}
can unfortunately not be expressed in a closed form which is valid for all $\mu$ and $T$.
It is due to this fact, that the intuitively simple conclusions of this work may partially be obscured by the math involved.

We first review the scaling of the spectrum in the massless case. 
Secondly, we discuss the non--thermal (i.e.\ degenerate) case in which the temperature $T\rightarrow0$ and the chemical potential $\mu$ is the Fermi energy.
As an example for this case we consider a decoupled and completely degenerate gas of right--helical neutrinos in the early universe \cite{Chen:2015dka}
for which we derive the mean relic velocity.
Thirdly, we discuss the purely thermal case with general $T$ and $\mu = m$. 
Finally, we consider the most general case with no constraints on $T$ and $\mu$.
We also compute the relic phase space densities for all cases. 

In the course of this work we present a new analytical expansion of the Fermi--Dirac integral for the case $T\rightarrow0$, $\mu\rightarrow m$, and an improved expansions for the case
$T\gg m$ and/or $\mu\gg m$ which is for our purpose superior to the Sommerfeld expansion. For the case $T=0$ all results are exact.

\section{Scaling of the spectrum}
\subsection{Massless case}

Let us briefly recall how one arrives at the insight that the phase space distribution of massless particles is invariant under spatial expansion, 
and then clarify how this statement fails to be true if particles have non--zero mass. 

The energy spectrum of fermions with mass $m$ and chemical potential $\mu$ at a temperature $T$ is given by \eqref{eq:FermiSpectrum}.
Setting $\mu=m=0$, the density of particles is calculated by integrating the spectrum over the physical range. It is given by
\begin{equation}\label{eq:Masslessdensity}
 n(T,\mu=0)~=~\int\displaylimits_{m=0}^\infty\!\mathrm{d}E\,\frac{\mathrm{d}n}{\mathrm{d}E}~=~\frac{g}{2\pi^2}\,\frac{3}{2}\,\zeta(3)\,T^3\;.
\end{equation}
Let us now assume that this gas is instantaneously decoupled from the thermal bath. 
The volume $V$ that the particles occupy, however, is assumed to evolve according to
\begin{equation}
 V_0~\rightarrow~V~=~V_0\,\frac{R^3}{R_0^3}\;,
\end{equation}
where $R$ is the so--called scale factor. Zeros denote the corresponding quantity at decoupling throughout this work.

Assuming that the number of particles is conserved, the density of particles has to scale inversely proportional to the volume,
\begin{equation}
 n
~=~n(T_0,\mu_0) \frac{R_0^3}{R^3}\;.
\end{equation}
The crucial assumption of this work is that we can -- even after decoupling, still -- describe the density as one 
of the two--parameter family of functions $n=n(T,\mu)$.\footnote{%
This assumption is automatically met by a massless gas and a degenerate massive Fermi gas. 
However, it is generally not met by a non--interacting massive gas. Assuming that $n=n(T,\mu)$ holds for a decoupled massive gas implicitly 
contains the assumption of the presence of sufficiently strong (elastic) self--interactions such that an equilibrium spectrum is retained.}
The intersection of the two functions $n(T,\mu)$ and $n(T_0,\mu_0)\times(R_0^3/R^3)$ then implicitly defines the functions $T(R)$ and $\mu(R)$ by
\begin{equation}\label{eq:CentralCondition}
n(T(R),\mu(R))~:=~n(T_0,\mu_0) \frac{R_0^3}{R^3}\;.
\end{equation}
Due to the uniqueness of this intersection, the functions $T(R)$ and $\mu(R)$ are unambiguously defined.
Using the exact result for the density \eqref{eq:Masslessdensity} on both sides of \eqref{eq:CentralCondition} 
we realize that the freely streaming massless gas, assumed $\mu_0=\mu=m=0$, behaves as if its temperature had been rescaled by 
\begin{equation}\label{eq:MasslessScaling}
T(R)~=~T_0\frac{R_0}{R}\;.
\end{equation}
Since the gas is decoupled, $T(R)$ is not a real temperature but the so--called ``effective'' temperature, which is used to describe the spectrum of 
non--interacting particles in the expanded volume. 

This establishes that a thermally decoupled and non--interacting gas of massless particles evolves under volume scaling 
as if one would have linearly rescaled its temperature. This fact (which is analogously derived for Bose statistics) 
is well established by the observed, flawlessly thermal spectrum of the Cosmic Microwave Background (CMB) (cf.\ e.g.\ \cite{Fixsen:2009}).

For clarity, let us also briefly mention the case that $m=0$ but $\mu\neq 0$. In this case, the density is given by
\begin{equation}\label{eq:MasslessdensityMu}
 n(T,\mu)~=~
 -\frac{g}{2\pi^2}\,2\,T^3\,\Li{3}{-\mathrm{e}^{\mu/T}}\;,
\end{equation}
where $\Li{s}{z}:=\sum^{\infty}_{k=1}\left(z^k/k^s\right)$ 
denotes the polylogarithmic function. In this case, the unique solution to \eqref{eq:CentralCondition} is given by the effective temperature and effective chemical potential
\begin{equation}\label{eq:TMuScalingMassless}
 T(R)~=~T_0\frac{R_0}{R}\;,~~~\text{and}~~~\mu(R)~=~\mu_0\frac{R_0}{R}\;.
\end{equation}
Therefore, there is only a little addendum to the above conclusion: A thermally decoupled gas of massless particles evolves under volume scaling 
as if one would have linearly rescaled its temperature \textit{and} its chemical potential.

In the massive case, the exact expression for the density of particles is more involved 
than \eqref{eq:Masslessdensity} or \eqref{eq:MasslessdensityMu} and this simple conclusion does not hold. 
Before we turn to the case of a massive quantum gas
there is one more pedagogical point which is crucial in order to understand this work.

Instead of working with scale factors, we may infer the (relative) size of the universe by observing the spectrum of a decoupled massless species, 
such as photons of the CMB. That is, we use the temperature of the CMB, $T_\gamma$, as a measure to gauge the (relative) size of the universe via \eqref{eq:MasslessScaling}.
As usual, we assume entropy conservation in the thermal bath. Any reheating of the CMB, happening in between the decoupling of the massive quantum gas and today,
then can be taken into account by rescaling the CMB temperature by a factor $[g_{*S}^\mathrm{today}/g_{*S}^{\mathrm{dec}}]^{1/3}$. Here $g_{*S}^\mathrm{today(dec)}$ stands for the effective number of massless degrees of freedom in entropy today or at the time when the 
massive gas decouples from the thermal bath, respectively. The ratio of scale factors, therefore, can be expressed as 
\begin{equation}\label{eq:TSdefinition}
\frac{\TS}{T_{0}}~:=~\frac{R_0}{R}~=~\left(\frac{g_{*S}(T_\gamma)}{g_{*S}(T_{0})}\right)^{1/3}\frac{T_\gamma}{T_{0}}\;.
\end{equation}
Here we have defined the scaling temperature $\TS$, which we will use throughout this work to characterize the volume scaling of the universe.
For example, assuming that a species has been in thermal equilibrium with the CMB until a decoupling temperature $T_0$ at which it had a chemical potential $\mu_0$, 
and that it is freely streaming after, the density at a later time is given by 
\begin{equation}\label{eq:scaledDensity}
 n(\TS)~=~n(T_0,\mu_0) \frac{\TS^3}{T_0^3}\;.
\end{equation}
The requirement $n(\TS)=n(T(\TS),\mu(T_S))$ then implicitly defines the effective temperature $T(\TS)$ and the effective chemical potential $\mu(T_S)$.
Note that in the massless case (and only there) the effective temperature coincides with the scaling temperature.

One should also remark that the initial temperature of the massive gas at decoupling can, in principle, be freely chosen and does not need to coincide with the temperature of the massless gas
which we use to characterize the scaling of the volume. 
This point is mostly relevant for the degenerate gas in the following section but may also be of interest for the treatment of distortions of the initial spectrum \cite{Dolgov:2002wy, Mangano:2005cc}, 
and more generally also for sectors which have not been in thermal contact with the standard model after all \cite{Bernal:2015ova}.

Let us now investigate the evolution of the spectrum of a massive Fermi gas. For the ease of the discussion we will first treat a degenerate ($T=0$) Fermi gas,
then the case of a minimal chemical potential $\mu=m$, and tackle the most general case in the end.

\subsection{Non--thermal case $\boldsymbol{(T=0)}$}

In the limit $T\rightarrow0$ the energy spectrum \eqref{eq:FermiSpectrum} reduces to 
\begin{equation}
 \frac{\mathrm{d}n}{\mathrm{d}E}~=~\frac{g}{2\pi^2}\,E\left(E^2-m^2\right)^{1/2}\,\Theta(\mu-E)\;,
\end{equation}
which can be integrated exactly resulting in
\begin{equation}\label{eq:NonThDensity}
 n(T=0,\mu)~=~\int\displaylimits_m^\infty\mathrm{d}E\,\frac{\mathrm{d}n}{\mathrm{d}E}~=~\frac{g}{6\pi^2}\,\left(\mu^2-m^2\right)^{3/2}\;.
\end{equation}
This is in agreement with \cite{Kolb:1990vq} where terms of the order $\mathcal{O}(m/\mu)$ have been neglected.
Restricting the density to be a real quantity we see from \eqref{eq:NonThDensity} that $\mu$ can only take values in the interval $[m,\infty)$. This is in agreement with the interpretation
of $\mu$ being the chemical potential, i.e.\ the energy needed in order to add one additional particle to the system.

\begin{figure}[t]
\includegraphics[width=0.5\textwidth]{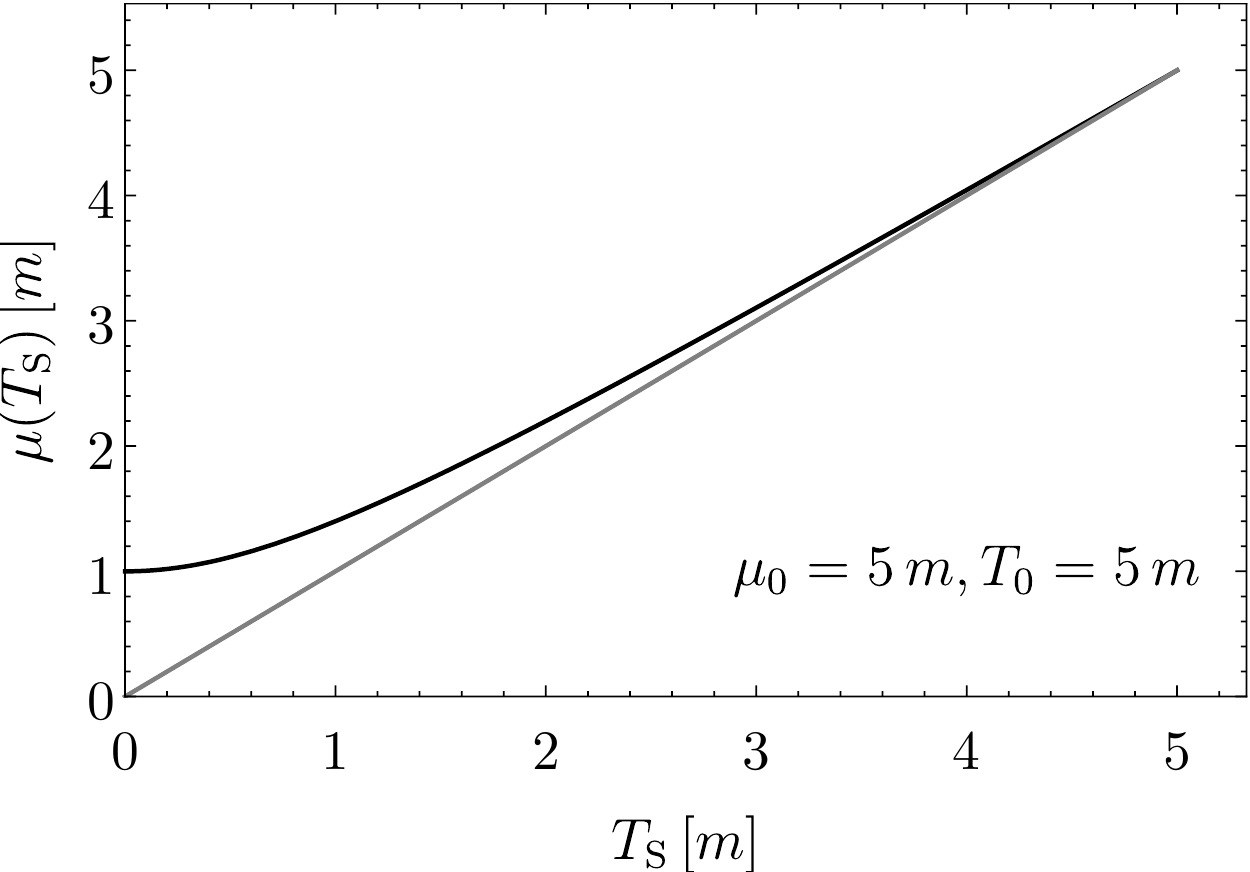}
\caption{Fermi energy of a degenerate massive Fermi gas in dependence of the volume scaling, here expressed in terms of $\TS$. 
For comparison we also show the scaling in the massless case (gray).}
\label{fig:MuTs}
\end{figure}

\begin{figure}[t]
\begin{minipage}[t]{0.5\linewidth} 
\includegraphics[width=0.9\textwidth]{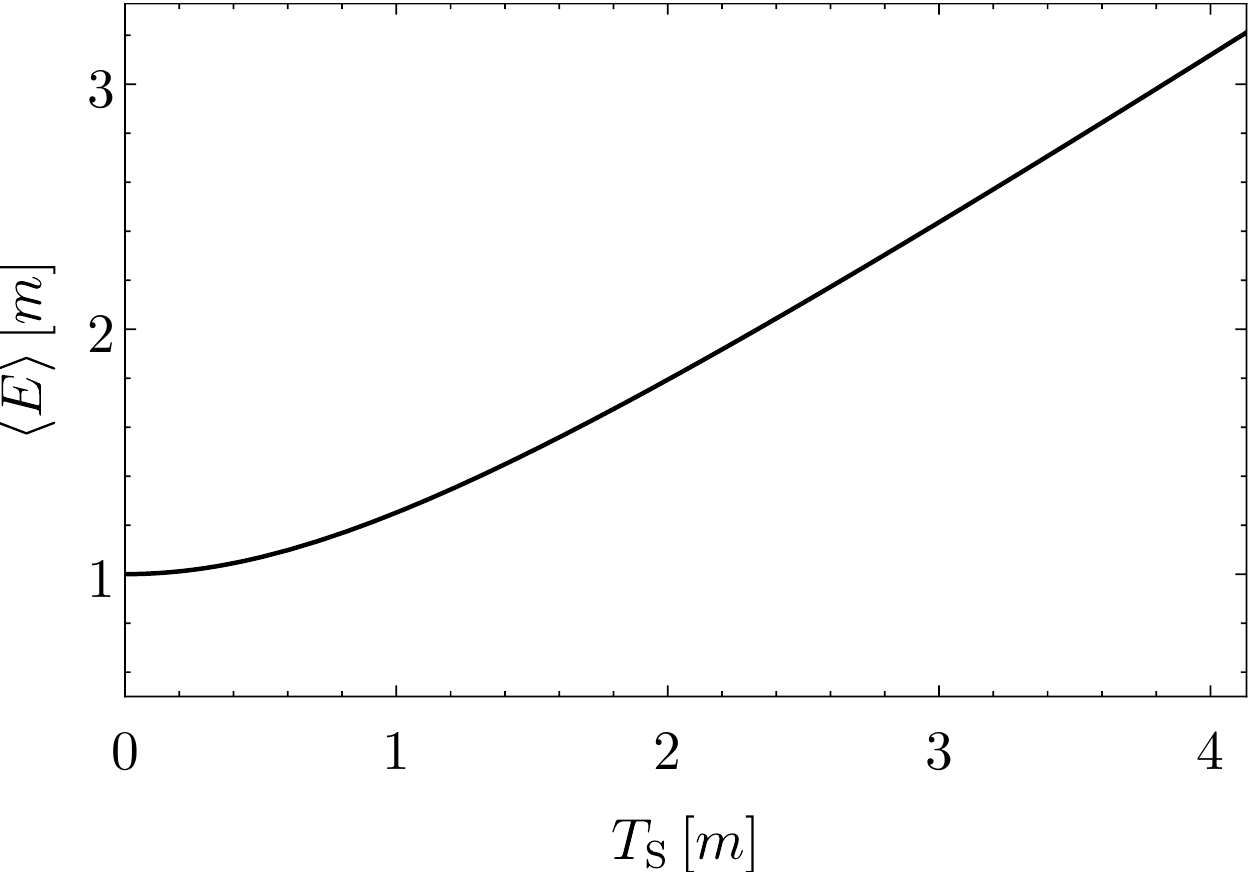}
\end{minipage}%
\begin{minipage}[t]{0.5\linewidth} 
\includegraphics[width=0.9\textwidth]{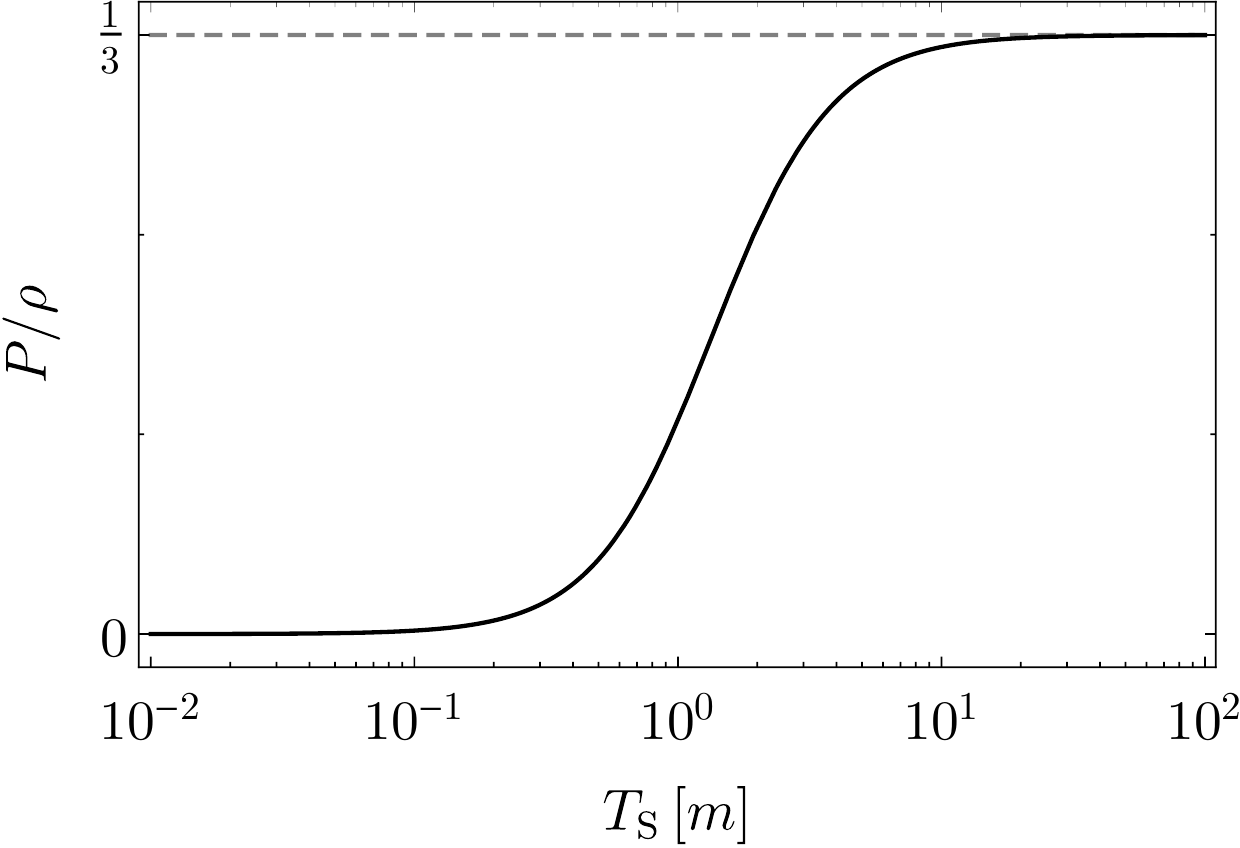}
\end{minipage}
\caption{Mean energy per particle (left) and the equation of state (right) in dependence of the volume scaling, here expressed in form of the scaling temperature $T_S$.
Initial parameters have been chosen to be $(T_0,\mu_0)=(5,5)$.}
\label{meanEEOS}
\end{figure}

Our goal is to infer the spectrum of a freely expanding -- yet completely degenerate -- gas in dependence of the volume expansion of the universe.
Since the spectrum is characterized by one parameter, namely $\mu$, we will be able to express the form of the spectrum via a scaling dependent $\mu(\TS)$. 
As initial conditions we assume that there exists a density of completely decoupled and degenerate fermions $n(\mu_0)$ with an initial Fermi energy $\mu_0$ in a given initial volume. 
The initial size of the volume is characterized by the initial background temperature $T_0$, and the expanded volume will be characterized by $\TS$. 
Typically $\mu_0\gg m$ but we do not require this here. As a consequence of particle number conservation the density at later times is given by
\begin{equation}\label{eq:densityOfT}
 n(\TS)~=~n(0,\mu_0)\left(\frac{\TS}{T_0}\right)^3~=~\frac{g}{6\pi^2}\,\left(\frac{\mu_0^2-m^2}{T_0^2}\right)^{3/2}\TS^3\;.
\end{equation}
We emphasize that $\mu_0$ and $T_0$ are just parameters of this setting, the only dynamical variable is \TS.
On the other hand, we can implicitly define a scaling temperature dependent Fermi energy via
\begin{equation}\label{eq:densityOfMu}
  n(\TS)~=:~n(0,\mu(\TS))~=~\frac{g}{6\pi^2}\,\left[\mu(\TS)^2-m^{2}\right]^{3/2}\;.
\end{equation}
Comparing \eqref{eq:densityOfT} and \eqref{eq:densityOfMu} we can extract the scaling dependent Fermi energy as a function of the scaling temperature $\TS$ and find
\begin{equation}\label{eq:muOfT}
 \mu(\TS)~=~\sqrt{m^2+\left(\frac{\mu_0^2-m^2}{T_0^2}\right) \TS^2}\;.
\end{equation}
The important result from this exact discussion is \eqref{eq:muOfT}, cf.\ \Figref{fig:MuTs}.
This relation describes the change of the spectrum in the expanding universe, i.e.\ how the Fermi energy decreases as the volume scales up. 
We note that the scaling relation \eqref{eq:TMuScalingMassless} for the massless case,
i.e.\ $\mu(\TS)\propto \TS\propto1/R$, is only approximately valid for scaling temperatures $\TS\gg m$. At low scaling temperatures, corresponding to large volume or late times,
the scaling of $\mu(\TS)$ is modified. Physically, this reflects the fact that the mean energy per particle cannot be reduced below $m$ (\Figref{meanEEOS}).

We can use \eqref{eq:muOfT} in order to obtain the correct scaling behavior of all quantities that one can derive from the spectrum. 
In order to keep the presentation of results simple, we will use the reduced scaling temperature $\TSEff$ which is defined by
\begin{equation}
 \TSEff~:=~\left(\frac{\mu_0^2-m^2}{T_0^2}\right)^{1/2}\TS\;.
\end{equation}
For a completely degenerate Fermi gas exact expressions for the energy density and pressure can be obtained, they are given by
\begin{align}\label{eq:NonThRho}
 \rho(\mu)~&=~\frac{1}{8}\,\mu\left(\mu^2-m^2\right)^{1/2}\left(2\mu^2-m^2\right)+m^4 \ln \frac{m}{\mu+\sqrt{\mu^2-m^2}}\;,\\ \label{eq:NonThP}
 P(\mu)~&=~\frac{1}{8}\,\mu\left(\mu^2-m^2\right)^{1/2}\left(2\mu^2-5m^2\right)-3\,m^4 \ln \frac{m}{\mu+\sqrt{\mu^2-m^2}}\;.
\end{align}
The important point is that we can now use \eqref{eq:muOfT} in order to obtain $\rho$, $P$, and also the equation of state $\omega=P/\rho$ in dependence 
of the volume scaling. Using \eqref{eq:NonThRho}, \eqref{eq:NonThP}, and \eqref{eq:muOfT}, the equation of state in dependence of the scaling temperature is given by (cf.\ \Figref{meanEEOS})
\begin{equation}
 P/\rho\,(\TS)~=~\frac{\left(\frac{\TSEff}{m}\right)\left[1+\left(\frac{\TSEff}{m}\right)^2\right]^{1/2}\left[\frac{2}{3}\left(\frac{\TSEff}{m}\right)^2-1\right]+\mathrm{arcsinh}\left(\frac{\TSEff}{m}\right)}
 {\left(\frac{\TSEff}{m}\right)\left[1+\left(\frac{\TSEff}{m}\right)^2\right]^{1/2}\left[2\left(\frac{\TSEff}{m}\right)^2+1\right]-\mathrm{arcsinh}\left(\frac{\TSEff}{m}\right)}\;.
\end{equation}
This function smoothly interpolates between the two limiting cases 
\begin{equation}\label{eq:EoSLimits}
 P/\rho\,(\TS\rightarrow\infty)~=~\frac{1}{3}\qquad\text{and}\qquad P/\rho\,(\TS\rightarrow0)~=~0\,,
\end{equation}
and describes the relativistic to non--relativistic transition of the non--interacting degenerate massive Fermi gas in an expanding volume.
In complete analogy, the mean energy per particle $\rho/n$ (cf.\ \Figref{meanEEOS}) can be computed from \eqref{eq:NonThDensity} and \eqref{eq:NonThRho}, and its scaling
dependent value can be obtained by performing the formal replacement $\mu\rightarrow\mu(\TS)$. 

The knowledge of how the spectrum changes with the volume scaling allows us to compute also other quantities with full scaling dependence.
For the gravitational clustering of relic neutrinos, for example, an important input is their mean velocity \cite{Singh:2002de, Ringwald:2004np}. The mean value of some quantity $X$ with respect to a given spectrum is defined by
\begin{equation}
 \langle X \rangle~:=~\frac{1}{n}\,\int\displaylimits_m^\infty\mathrm{d}E\,\frac{\mathrm{d}n}{\mathrm{d}E}\,X\;,
\end{equation}
where we have normalized to the density. The velocity is given by $v=|\partial E/\partial p|$, where $E=\sqrt{p^2+m^2}$.
For the non--thermal ($T=0$) spectrum the mean velocity can be computed to be
\begin{equation}
 \langle v (\mu)\rangle ~=~ \frac{\left(\mu-m\right)^2\left(\mu+2m\right)}{\left(\mu^2-m^2\right)^{3/2}}\;.
\end{equation}
By plugging in the scaling temperature dependence of $\mu$ as given by \eqref{eq:muOfT}, we can obtain the scaling temperature dependence of the mean velocity.
It is given by
\begin{equation}
 \langle v (\TS)\rangle ~=~ \left(\frac{m}{\TSEff}\right)^3\left\{ 1-\left[1+\left(\frac{\TSEff}{m}\right)^2\right]^{1/2}\right\}\left\{ 2+\left[1+\left(\frac{\TSEff}{m}\right)^2\right]^{1/2}\right\}\;.
 \end{equation}
This result is exact. We can expand it for scaling temperatures $\TS\ll m$, i.e.\ at late times in the evolution of the universe, and find
\begin{equation}
 \langle v(\TS)\rangle~=~\frac{3}{4}\left(\frac{\mu_0^2-m^2}{T_0^2}\right)^{1/2}\frac{\TS}{m}+\mathcal{O}\left[\left(\frac{\TS}{m}\right)^3\right]\;.
\end{equation}
We can use this result in order to compute the mean velocity of the recently proposed non--thermal (degenerate) background of RH neutrinos \cite{Chen:2015dka}.
In that case, $m\lll T_0$, and $\mu_0/T_0$ is bounded above by observations of the energy density during BBN, typically expressed in terms of $\Delta N_{\mathrm{eff}}$ (cf.\ e.g.\ \cite{Ade:2015xua}).
The mean velocity of non--thermal relic neutrinos at late times then is given by 
\begin{equation}
\langle v_\text{C$\nu$B, non--thermal} \rangle~=~572 \left( 1+\mathsf{z} \right)
\left(\frac{0.1\,\ev}{m_{\nu}}\right)\left(\frac{\Delta N^{\mathrm{nt}}_{\mathrm{eff}}}{0.7} \right)^{1/4} \text{km}\,\text{s}^{-1}\;.
\end{equation}
We have expressed this in the conventional form where $\mathsf{z}$ denotes the red--shift relative to today and 
$\Delta N^{\mathrm{nt}}_{\mathrm{eff}}$ is the additional number of effective degrees of freedom during BBN caused by 
the non--thermal relic neutrinos.

\subsection{Thermal case $\boldsymbol{(\mu=m)}$}

\begin{figure}[t]
\subfigure[]{
\includegraphics[width=0.31\textwidth]{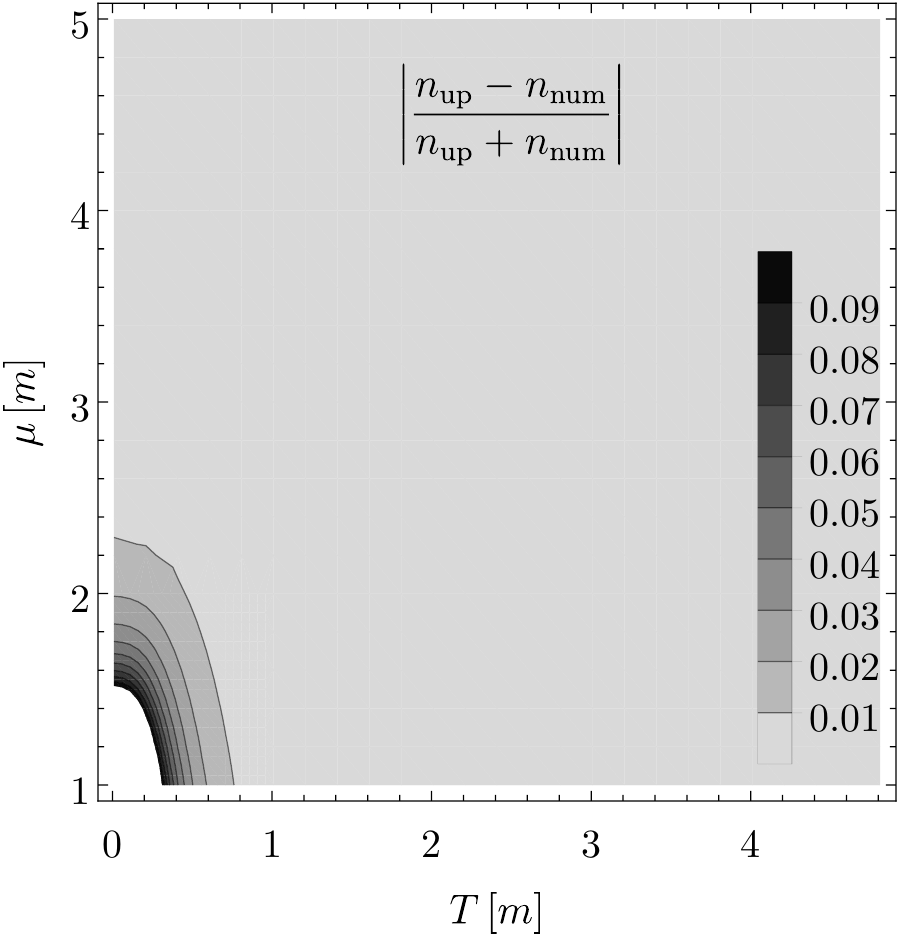}
}
\subfigure[]{
\includegraphics[width=0.31\textwidth]{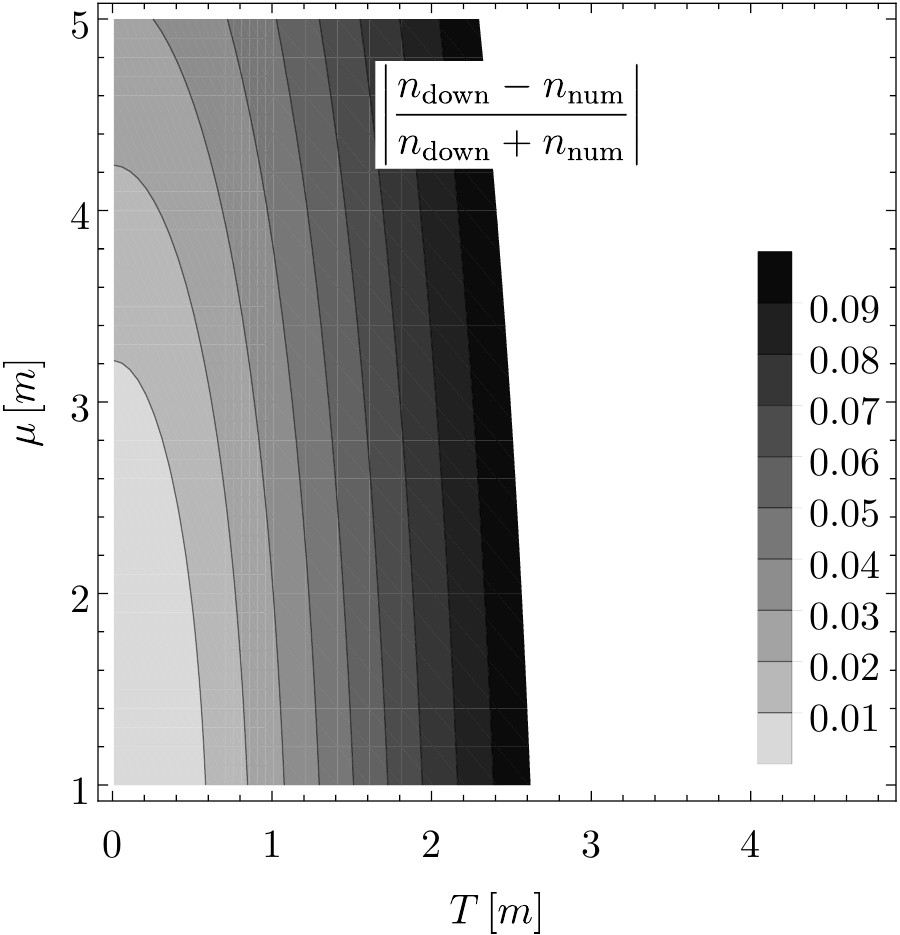}
}
\subfigure[]{
\includegraphics[width=0.31\textwidth]{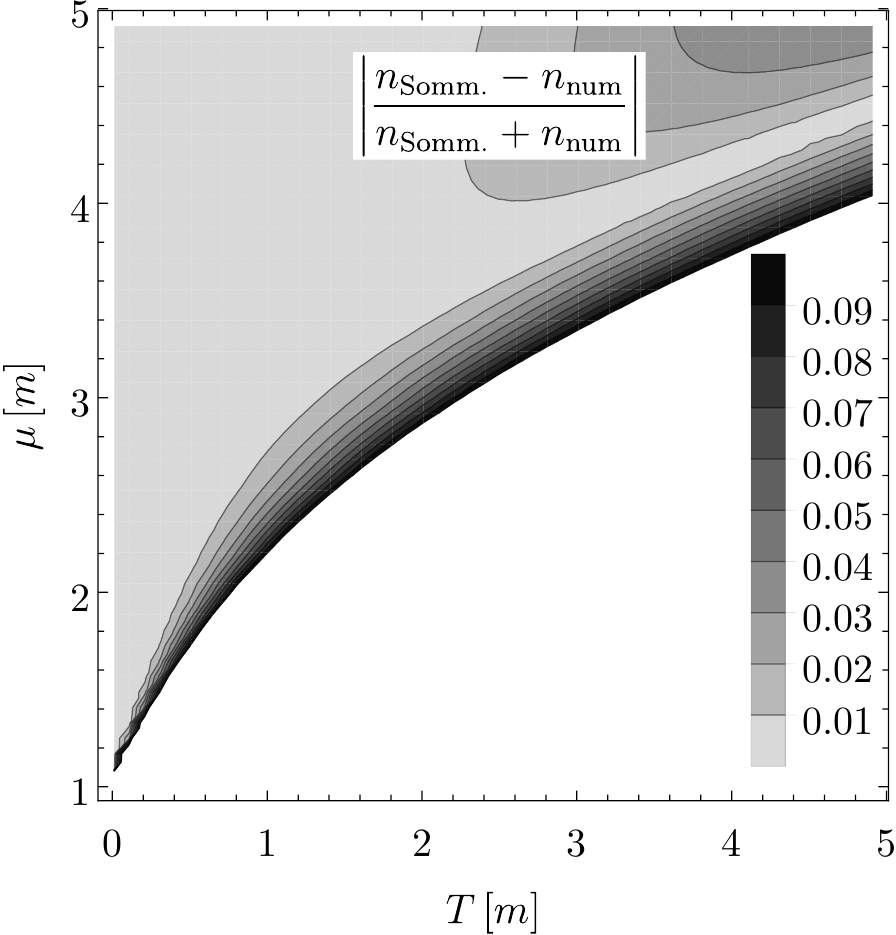}
}
\caption{Deviations from the numerically computed density $n(T,\mu)$ relative to our approximation ``up'' \eqref{eq:ExactDensityUp} (a), ``down'' \eqref{eq:ExactDensityDown} (b), and the Sommerfeld approximation (c).}
\label{fig:nRelErrors}
\end{figure}

Let us now consider the case in which $T>0$. 
We will make use of two different approximations for the 
Fermi--Dirac integral \eqref{eq:FermiDiracIntegral}. 
The first one -- in the following referred to as ``up'' -- is valid at high temperatures $T\gg m$ and/or high chemical potentials $\mu\gg m$; 
the second one -- ``down'' -- is valid in the case $T\gtrsim0$ and $\mu\gtrsim m$ (\Figref{fig:nRelErrors}). 
The ``up'' approximation will be used to compute the initial density of the massive Fermi gas in the hot early universe. 
After the decoupling, the density will scale down inversely proportional to the volume until today.
At low scaling temperatures $\TS\ll m$, the scaled density has to match the density which can be calculated in the ``down'' approximation.
This is true because even though $T\propto \TS$ does not hold at low temperatures, we still know the limiting behavior $T(\TS\rightarrow0)\rightarrow0$ 
simply because the particle density has to vanish for an infinitely upscaled volume. The goal is to extract the leading order behavior of $T(\TS)$ in the limit of high and low scaling 
temperatures. We will then present a phenomenological interpolation function for $T(\TS)$ which fulfills the 
limiting behavior at low and high $\TS$ and correctly reproduces the scaled density.

An excellent approximation for the density in case that $T\gg m$ and/or $\mu\gg m$ is given by 
\begin{subequations}
\begin{equation}\label{eq:ExactDensityUp}
 n(T,\mu)_{\mathrm{up}}\,\frac{2\pi^2}{g}~=~-2\,T^3\left[\Li{3}{-z}+\frac{m}{T}\Li{2}{-z}-\frac{1}{4}\frac{m^2}{T^2}\log(1+z)+\frac{1}{12}\frac{m^3}{T^3}\frac{z}{1+z}\right]\;,
\end{equation}
where we have used $z:=\exp({\frac{\mu-m}{T}})$, which should not be confused with the red--shift above. In the opposite case,
namely $T\ll m$ and $\mu\gtrsim m$, a very good approximation for the density is given by
\begin{equation}\label{eq:ExactDensityDown}
 n(T,\mu)_{\mathrm{down}}\,\frac{2\pi^2}{g}~=~-\sqrt{\frac{\pi}{2}}(m\,T)^{3/2}\left[\Li{3/2}{-z}+\frac{15}{8}\frac{T}{m}\Li{5/2}{-z}+\frac{105}{128}\frac{T^2}{m^2}\Li{7/2}{-z}\right]\;.
\end{equation}
\end{subequations}
These results stem from a novel asymptotic series expansion of the Fermi--Dirac integral \eqref{eq:FermiDiracIntegral} which can be obtained by expanding the integrand in the limits $E\gg 2m$ and $E\ll 2m$.
The general form of these series is given in Appendix \ref{app:FermiDiracSeries}. Comparing our approximations to the numerically integrated density\footnote{%
All numerical computations in this work have been done with \textsc{mathematica}.} 
we see that we can reliably compute the density 
within these approximations (\Figref{fig:nRelErrors}). 
Combining our approximations, the relative error to the numerically evaluated integral is at maximum $1\%$ in the region around $T\sim m$, and negligible everywhere else. 
For comparison, we also show the density as obtained with the well--known approximation by Sommerfeld \cite{Sommerfeld:1928,Ashcroft} which holds only for the case that $\mu> m$ and $T\ll\mu$.\footnote{%
The Sommerfeld approximation requires the function in the numerator of \eqref{eq:FermiSpectrum} to be sufficiently smooth around $E=\mu$ such that one can Taylor expand it around $\mu$ 
(cf.\ e.g.\ \cite{Ashcroft}). The density spectrum of \eqref{eq:FermiSpectrum} does not fulfill this criterion in the limit $\mu\rightarrow m$. This is the reason why the approximation fails 
to reliably compute the density in the limit of low scaling temperatures corresponding to a large upscaled volume.}

\begin{figure}[t]
\includegraphics[width=0.5\textwidth]{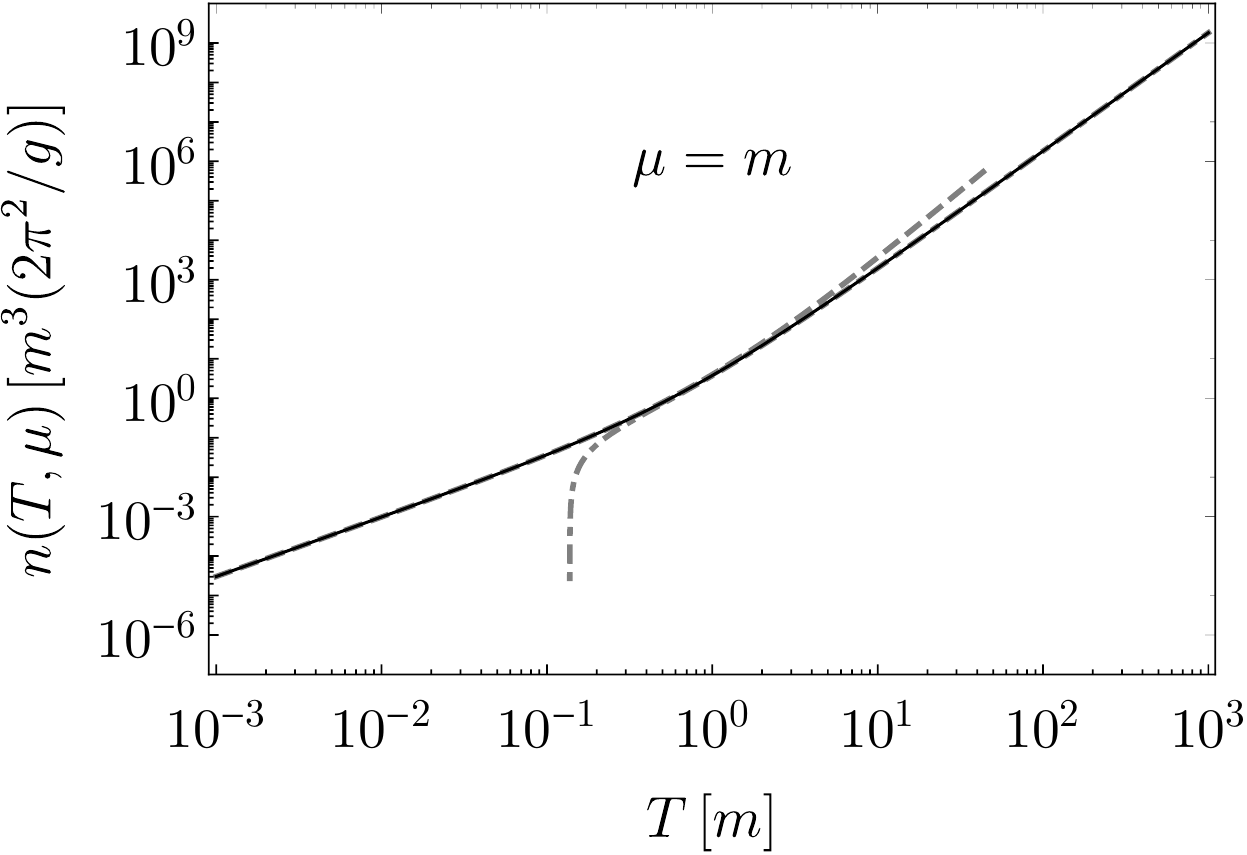}
\caption{Density $n(T,\mu=m)$ as calculated with our approximations ``up'' \eqref{eq:DensityUp} (dot--dashed) and ``down'' \eqref{eq:DensityDown} (dashed) 
compared to the numerically integrated density (black, solid).}
\label{fig:DensityMuM}
\end{figure}

Let us now focus on the case that $\mu=m$ exactly. Therefore, $z=1$ and eqs.\ \eqref{eq:ExactDensityUp} and \eqref{eq:ExactDensityDown} reduce to
\begin{subequations}
\begin{align}\label{eq:DensityUp}
 n(T,m)_\mathrm{up}\,\frac{2\pi^2}{g}~=&~\frac{3}{2}\,\zeta(3)\,T^3+\frac{\pi^2}{6}\,T^2\,m+\frac{\ln(2)}{2}\,T\,m^2-\frac{m^3}{12}\,,\quad\text{and}\\ \label{eq:DensityDown}
 n(T,m)_\mathrm{down}\,\frac{2\pi^2}{g}~=&~\sqrt{\frac{\pi}{2}}(m\,T)^{3/2}\left[\left(1-\frac{1}{\sqrt{2}}\right)\zeta(3/2)+\mathcal{O}\left(\frac{T}{m}\right)\right]\;,
\end{align}
\end{subequations}
respectively. Note the familiar leading order term of \eqref{eq:DensityUp}.
In the ultrarelativistic or massless
case, corresponding to $m\rightarrow0$, this is the only non--vanishing term of both expansions.\footnote{%
The leading order term of \eqref{eq:DensityDown} agrees with the ``non--relativistic'' approximation in the literature (cf.\ e.g.\ \cite[eq.\ (3.55)]{Kolb:1990vq}, taking $\mu\to m$)
only up to the factor $(1-2^{-1/2})\zeta(3/2)\approx0.77$ which is often neglected.}
The density obtained with equations \eqref{eq:DensityUp} and \eqref{eq:DensityDown} next to a numerical computation is shown in \Figref{fig:DensityMuM}.

Consider now a gas which has $\mu_0=\mu=m$ and is decoupled at a temperature $T_0\gg m$. The density at later times is given by \eqref{eq:scaledDensity}, where we use that 
\begin{equation}
 n(T_0,m)~\approx~n_\mathrm{up}(T_0,m)\;.
\end{equation}
Completely analogous to the non--thermal case, we want to find a function $T(\TS)$ which describes the spectrum after decoupling, under the crucial assumption that 
the total particle number is conserved. This function is implicitly defined by the requirement
\begin{equation}\label{eq:DefiningTS}
 n(\TS)~=~n(T_0,m)\,\frac{\TS^3}{T_0^3}~=:~n(T(\TS),m)\;.
\end{equation}
The involved form of the expression for the density does, in contrast to the non--thermal case, not allow us to solve \eqref{eq:DefiningTS} for $T(\TS)$ in 
a closed form. We will, therefore, solve for $T(\TS)$ in form of a power series for high and low values of $\TS$ and then present a phenomenological fit function which interpolates 
between the two regimes. 

Close to decoupling, $T_0\gtrsim \TS\gg m$ and we are in the regime of high $\TS$. Therefore, we use $n_\mathrm{up}(T(\TS),m)$ on the right--hand side of \eqref{eq:DefiningTS} to approximate the scaled density. 
Expanding in a power series around $\TS=T_0$ and solving \eqref{eq:DefiningTS} order by order in $\TS$ we find
\begin{equation}\label{eq:TSExpansionUp}
 T(\TS\gg m)~\simeq~T_0+b^\mathrm{up}_1\,(\TS-T_0)+\frac{b^\mathrm{up}_2}{T_0}\,(\TS-T_0)^2+\dots\;.
\end{equation}
The expansion coefficients are given in Appendix \ref{app:TSupSeriesCoefficients}. 

\begin{figure}[t]
\includegraphics[width=0.5\textwidth]{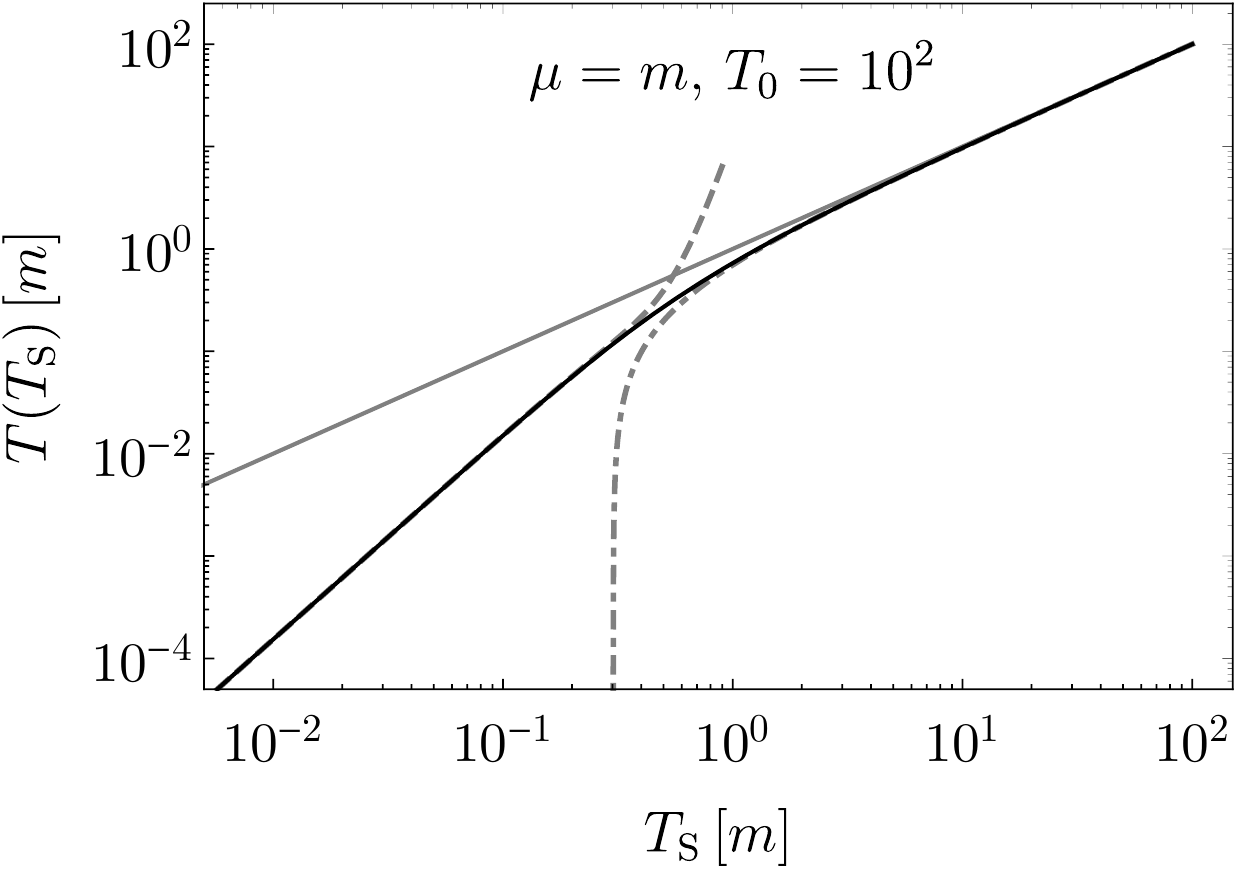}
\caption{Phenomenological fit function (black, solid) for the effective temperature $T(\TS)$ \eqref{eq:TTsfitfunction} as function of the scaling temperature as obtained from the series expansions for 
low (dashed) and high (dot--dashed) $\TS$. For comparison we also show the scaling in the massless case in which $T(\TS)=\TS$ (gray).}
\label{fig:TTs}
\end{figure}

At low $\TS$, we know that $T(\TS\rightarrow 0)\rightarrow 0$ and, therefore, use $n_\mathrm{down}(T(\TS),m)$ to approximate the scaled density on the right--hand side of \eqref{eq:DefiningTS}. 
We then expand in a power series around $\TS=0$ and solve \eqref{eq:DefiningTS} order by order in $\TS$ to find 
\begin{equation}\label{eq:TTsSeriesDown}
T(\TS\ll m)~\simeq~\frac{\xi\,n_0^{2/3}}{T_0^2} \frac{\TS^2}{m} + \left(\frac{\xi\,n_0^{2/3}}{T_0^2}\right)^2\,\frac{5}{8}\frac{(4-\sqrt{2})}{(2-\sqrt{2})}\frac{\zeta(5/2)}{\zeta(3/2)}\,\frac{\TS^4}{m^3} + \dots\;.
\end{equation}
Here we have used $n_0\equiv n(T_0,\mu_0)$ and $\xi$ which is defined by
\begin{equation}
 \xi~:=~\left(\frac{2}{\zeta(3/2)}\right)^{2/3}\,\left(\frac{3+2\sqrt{2}}{\pi}\right)^{1/3}~\approx~1.02832\;.
\end{equation}
In principle, one could expand $T(\TS)$ from above and below up to some order and match the two expansions somewhere around $\TS\sim m$. 
For practical purposes, however, it is much more convenient to have a complete analytic expression for $T(\TS)$ analogous to $\mu(\TS)$ above, cf.\ eq.\ \eqref{eq:muOfT}.
The requirements on such a function are: (i) It has to start at $\TS\gtrsim 0$ like $\TS^2$; (ii) It has to obey $T(\TS=T_0)=T_0$; (iii) For $\TS\rightarrow\infty$ it should raise linearly with $\TS$. 
A function which fulfills these requirements is given by
\begin{equation}\label{eq:TTsfitfunction}
T(\TS)~=~\frac{T_0}{2(\rho-1)}\,\left(-1+\sqrt{1+4\,\rho\,(\rho-1)\frac{\TS^2}{T_0^2}}\right)\;,
\end{equation}
where $\rho$ is, in principle, a free parameter. We can fix $\rho$ by comparing the density $n_\mathrm{down}(T(\TS),m)$
obtained with the low $\TS$ expansion of $T(\TS)$,
\begin{equation}\label{eq:lowTTs}
 T(\TS\ll T_0)~=~\rho\,T_0 \frac{\TS^2}{T^2_0}+\mathcal{O}\left(\frac{\TS^4}{T_0^4}\right)\;,
\end{equation}
to the correctly scaled density. 
In case that $\mu=m$, this implies that the first term of \eqref{eq:lowTTs} has to coincide with 
\eqref{eq:TTsSeriesDown} and we obtain $\rho=(\xi n_0^{2/3})/(mT_0)$. 
For the case $\mu=m$, the phenomenological function $T(\TS)$ as well as the two approximations for high and low $\TS$ are displayed in \Figref{fig:TTs}. 
Note that $n_\mathrm{down}$ is modified in the most general case ($\mu\neq m$) and so will be $\rho$, we will investigate this below.

\begin{figure}[t]
\includegraphics[width=0.5\textwidth]{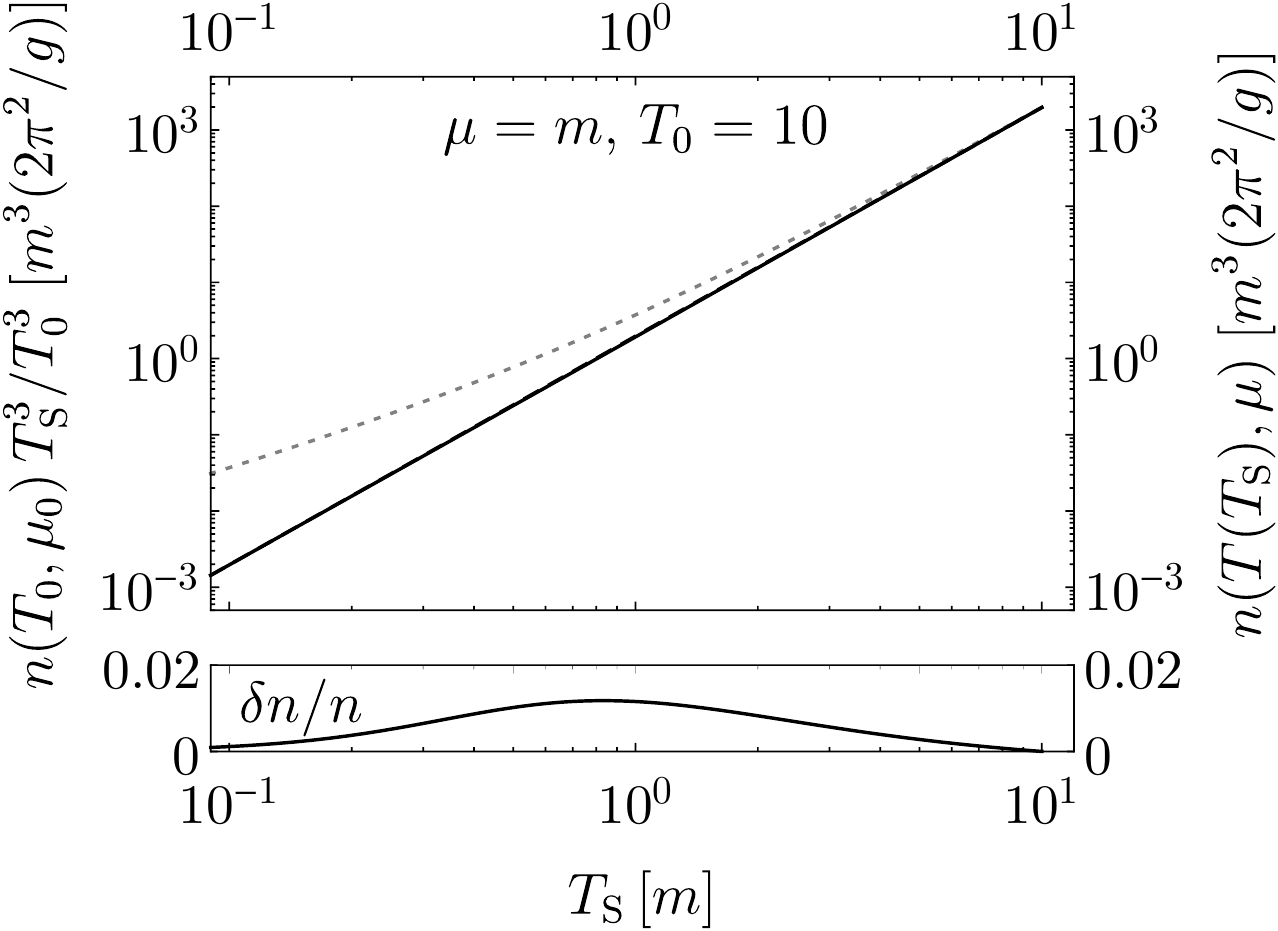}
\caption{The properly scaled density (black, dashed) compared to the density obtained by using the phenomenological function for the effective temperature \eqref{eq:TTsfitfunction} and numerical integration (black, solid).
The two curves overlap and below we show the relative error. 
In dotted gray we also show how the density would be overestimated with the assumption of linear scaling $T(\TS)=\TS$ in the massive case.}
\label{fig:ScaledDensityAndError}
\end{figure}

Let us also compare the correctly scaled density $n_0\times(\TS/T_0)^3$ to the density obtained from $n(T(\TS),m)$ with the phenomenological function for $T(\TS)$. 
The two densities are displayed in \Figref{fig:ScaledDensityAndError}, 
together with the relative error between the correctly scaled density and the scaled density obtained with the phenomenological fit function.
We see that the relative error is $\lesssim2\%$ around $\TS\sim m$ and lower at all other values.

The important result from this discussion is $T(\TS)$, cf.\ \Eqref{eq:TTsfitfunction} and \Figref{fig:TTs}. 
This function describes how the spectrum of a massive and self--interacting gas changes in the expanding universe, 
i.e.\ how the effective temperature decreases as the volume scales up. 
In particular we find that for $T_0\gg m$ the spectrum smoothly changes from a ``relativistic'' shape at decoupling to a ``non--relativistic'' shape at late times.
Note that this is the only possible behavior of an equilibrium spectrum in accordance with the conservation of particle number.

We want to stress that it is the effective temperature $T(\TS)$ and \textit{not} the scaling temperature $\TS$ which appears in the spectrum.
At late times $\TS\ll m$ the gas has long become non--relativistic and the effective temperature is given by
\begin{equation}
 T(\TS)~\simeq~\rho\,\frac{\TS^2}{T_0}~=~\frac{\xi\,n_0^{2/3}}{m\,T_0}\,\frac{\TS^2}{T_0}~=
~\left(\frac{3\,\zeta(3)}{\zeta(3/2)}\right)^{2/3}\,\left(\frac{3+2\sqrt{2}}{\pi}\right)^{1/3}\,\frac{\TS^2}{m}~\approx~1.52338\,\frac{\TS^2}{m}\;.
\end{equation}
For the last two steps we have assumed that $T_0\gg m$.
The final effective temperature, therefore, is independent of the initial conditions (given $T_0\gg m$) and only a function of the particle mass $m$ and the final scale factor $\TS$. 
Via \eqref{eq:TSdefinition}, $\TS$ can be expressed in terms of the CMB temperature $T_\gamma$, which enters as a measure of the relative size of the universe. 
This discussion applies to decoupled species which are non--relativistic in today's universe, which is the case whenever $m\gtrsim 10^{-4}\ev$.

\begin{figure}[t]
\subfigure[]{
\includegraphics[width=0.45\textwidth]{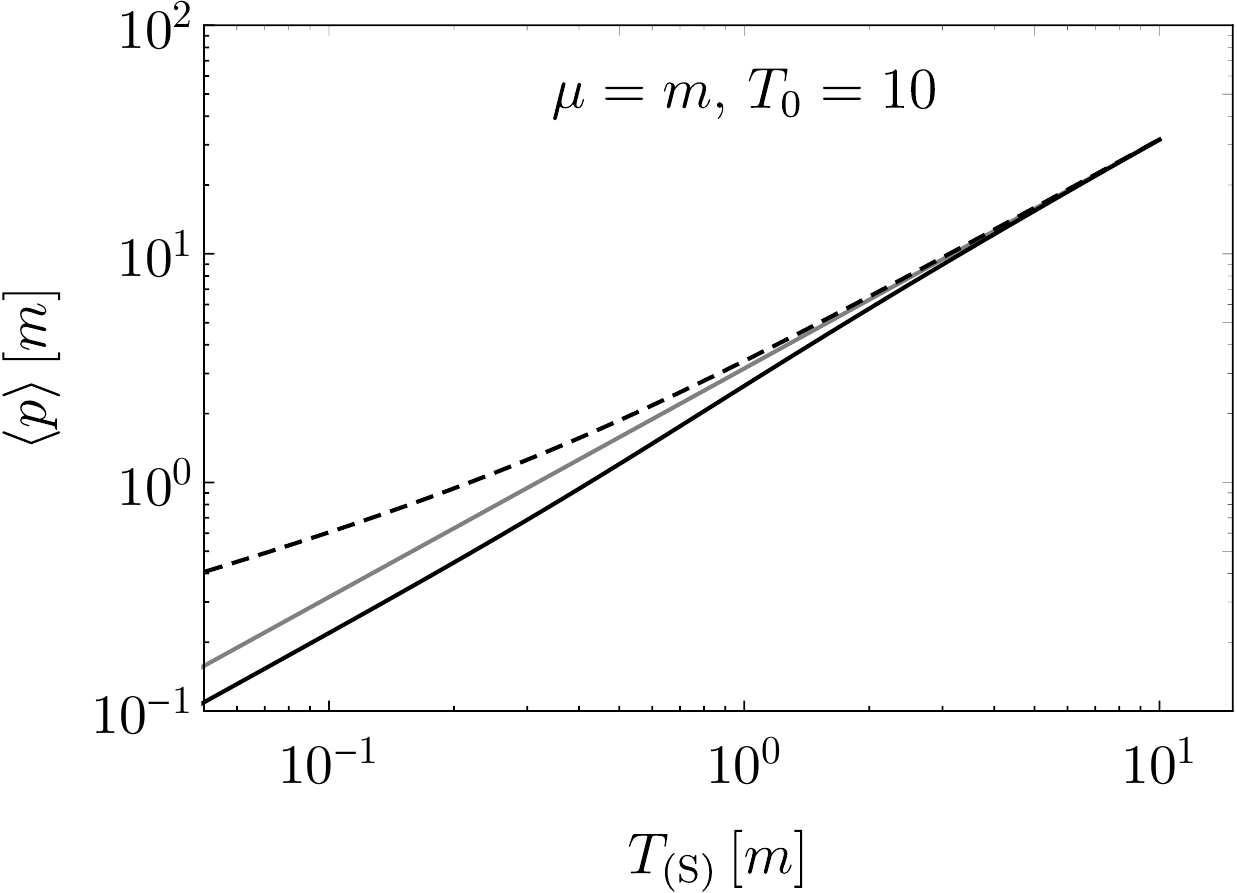}\label{fig:Meanp}
}
\subfigure[]{
\includegraphics[width=0.45\textwidth]{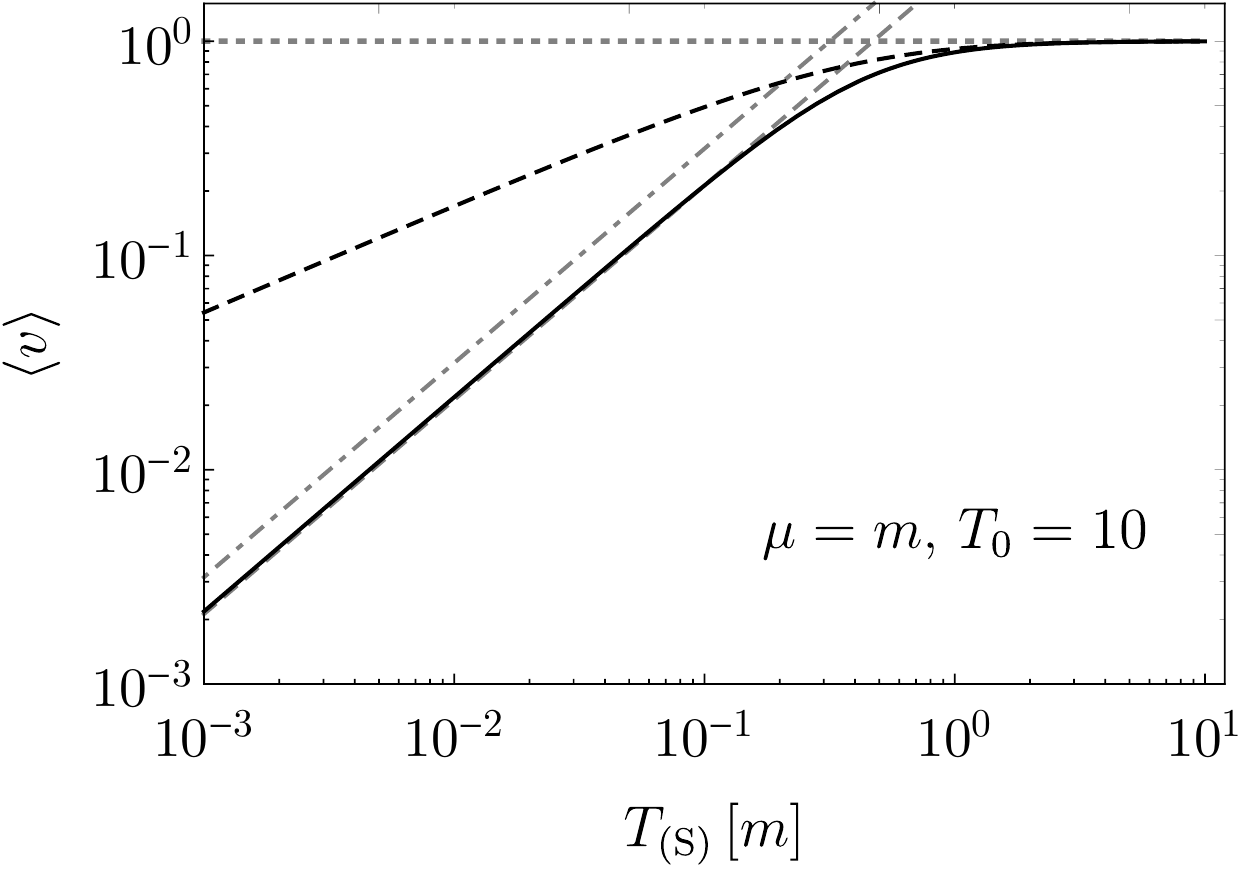}\label{fig:Meanv}
}
\caption{(a) Mean momentum in dependence of the real temperature $T$ (dashed) and in dependence of the volume scaling $\TS$ (black, solid).
In comparison we also show the massless case (gray) in which $T$ and $\TS$ coincide.
(b) Mean velocity for the real temperature case (dashed) and in dependence of the volume scaling (black, solid). We also show the two different linear approximations
$\langle v \rangle=3.15\,\TS/m$ (gray, dotdashed) and $\langle v \rangle=2.12\,\TS/m$ (gray, dashed).}
\end{figure}

Given $T(\TS)$, let us also compute the scaling dependence of some quantities which can be derived from the spectrum.
For example, take the mean momentum which for the spectrum \eqref{eq:FermiSpectrum} is given by
\begin{equation}
 \langle p(T,\mu)\rangle~=~-\frac{g}{2\pi^2}\,\frac{6\,T^4}{n(T,\mu)}\left(\Li{4}{-z}+\frac{m}{T}\,\Li{3}{-z}+\frac{m^3}{3\,T^3}\,\Li{2}{-z} \right)\;.
\end{equation}
The mean momentum for the real temperature case in comparison to the correctly scaled, i.e.\ scaling temperature dependent mean momentum (for $\mu=m$) is shown in \Figref{fig:Meanp}.
We also show the mean momentum for the massless case, where the real and scaling temperature dependent expressions for $\langle p \rangle$ coincide. 
Note that in the massive case $\langle p\rangle$ is not directly proportional to $\TS$ (or inverse proportional to $R$) for all regions, but only for high and low $\TS$ compared to the mass.

Let us also compute the mean velocity of relic fermions with the correct scaling dependence. 
The mean velocity for the spectrum \eqref{eq:FermiSpectrum} is given by
\begin{equation}
 \langle v(T,\mu)\rangle~=~-\frac{g}{2\pi^2}\,\frac{2\,T^3}{n(T,\mu)}\left(\Li{3}{-z}+\frac{m}{T}\,\Li{2}{-z} \right)\;.
\end{equation}
We are interested in the scaling dependent velocity after decoupling. 
Using $T(\TS)$ from \eqref{eq:TTsfitfunction} and the scaled density \eqref{eq:scaledDensity}, we expand $\langle v \rangle$ for small $\TS$ and obtain
\begin{equation}
 \langle v(\TS)\rangle~\simeq~\left(\frac{3}{2}\zeta(3)\right)^{1/3}\frac{\pi^2\,\xi^2}{6}\,\frac{\TS}{m}~\approx~2.12\,\frac{\TS}{m}\;.
\end{equation}

This value should be compared to the case of a massive gas with strictly no self--interactions such as the Standard Model cosmic neutrino background (C$\nu$B). 
In this case the statement of an effective temperature is not possible as the gas does not maintain an equilibrium spectrum after turning non--relativistic due to expansion. 
Nevertheless, a simple estimate for this case (based on footnote \ref{SMneutrinos}) yields
\begin{equation}
 \langle v \rangle~\simeq~
  3.15\,\frac{T_\nu}{m}\;,
\end{equation}
where $T_\nu$ is the \textit{scaling} temperature of the C$\nu$B.
This corresponds to a mean relic velocity
(cf.\ e.g.\ \cite{Bond:1980ha,Singh:2002de, Ringwald:2004np,Lesgourgues:2006nd, Lesgourgues:2014zoa,Safdi:2014rza})
\begin{equation}
\langle v_\text{C$\nu$B} \rangle~=~1.58\times10^{3} \left( 1+\mathsf{z} \right)
\left(\frac{0.1\,\ev}{m_{\nu}}\right)\text{km}\,\text{s}^{-1}\;.
\end{equation}
In contrast, if massive relic neutrinos would self--thermalize after their decoupling, 
they would obey a Fermi--Dirac spectrum with an effective temperature
\begin{equation}
 T_\nu^\mathrm{eff}~\approx~1.52338\,\frac{T_\nu^2}{m_\nu}~=~4.28\times10^{-7}\,\left(\frac{0.1\,\ev}{m_\nu}\right)\,\frac{\ev}{k_\mathrm{B}}~=~4.97\times10^{-3}\,\left(\frac{0.1\,\ev}{m_\nu}\right)\,\mathrm{K}\;,
 \end{equation}
which is valid for neutrino masses $m_\nu\gg T_\nu\approx 2\times10^{-4}\ev$. 
Their mean relic velocity would be suppressed by a factor $\sim2.12/3.15$.

In conclusion, we see that a decoupled but self--interacting massive species obeys a non--relativistic spectrum after it has been red--shifted below $\TS\sim m$,
implying that the mean relic velocity is suppressed compared to the strictly non--interacting case.

\subsection{General case}

So far, we have obtained the functions $T(\TS)$ and $\mu(\TS)$ only for the case that $\mu=m$ or $T=0$, respectively. 
Let us now discuss the most general case in which both, $T$ and $\mu$, are non--trivial. 
We seek functions $T(\TS)$ and $\mu(\TS)$ such as to fulfill
\begin{equation}
 n(T(\TS),\mu(\TS))~=~n(T_0,\mu_0)\,\frac{\TS^3}{T_0^3}\;.
\end{equation}
General arguments suggest that the functions $\mu(\TS)$ and $T(\TS)$ keep approximately the same functional form as in \eqref{eq:muOfT} and \eqref{eq:TTsfitfunction}.
As the scaled spectrum of particles has to match the thermal spectrum of particles at decoupling, the requirements $\mu(\TS=T_0)=\mu_0$ and
$T(\TS=T_0)=T_0$ must be fulfilled. Furthermore, since the density must still vanish in the infinite volume case, we know that
$T(\TS\rightarrow0)\rightarrow0$ and $\mu(\TS\rightarrow0)\rightarrow m$. The scaling of the density implies that both functions should start quadratically
in $\TS$ at low $\TS$. This brings us back to the functions $\mu(\TS)$ and $T(\TS)$ as obtained above.

\begin{figure}[t]
\subfigure[]{
\includegraphics[width=0.4\textwidth]{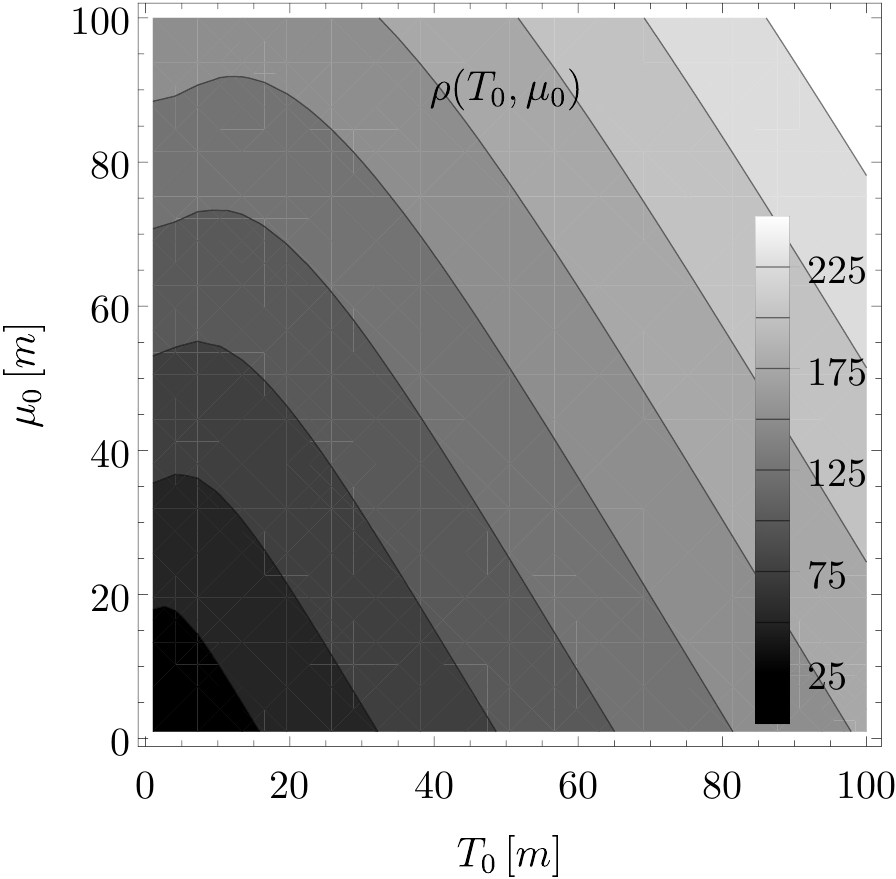}\label{fig:RhoValues}
}
\subfigure[]{
\includegraphics[width=0.5\textwidth]{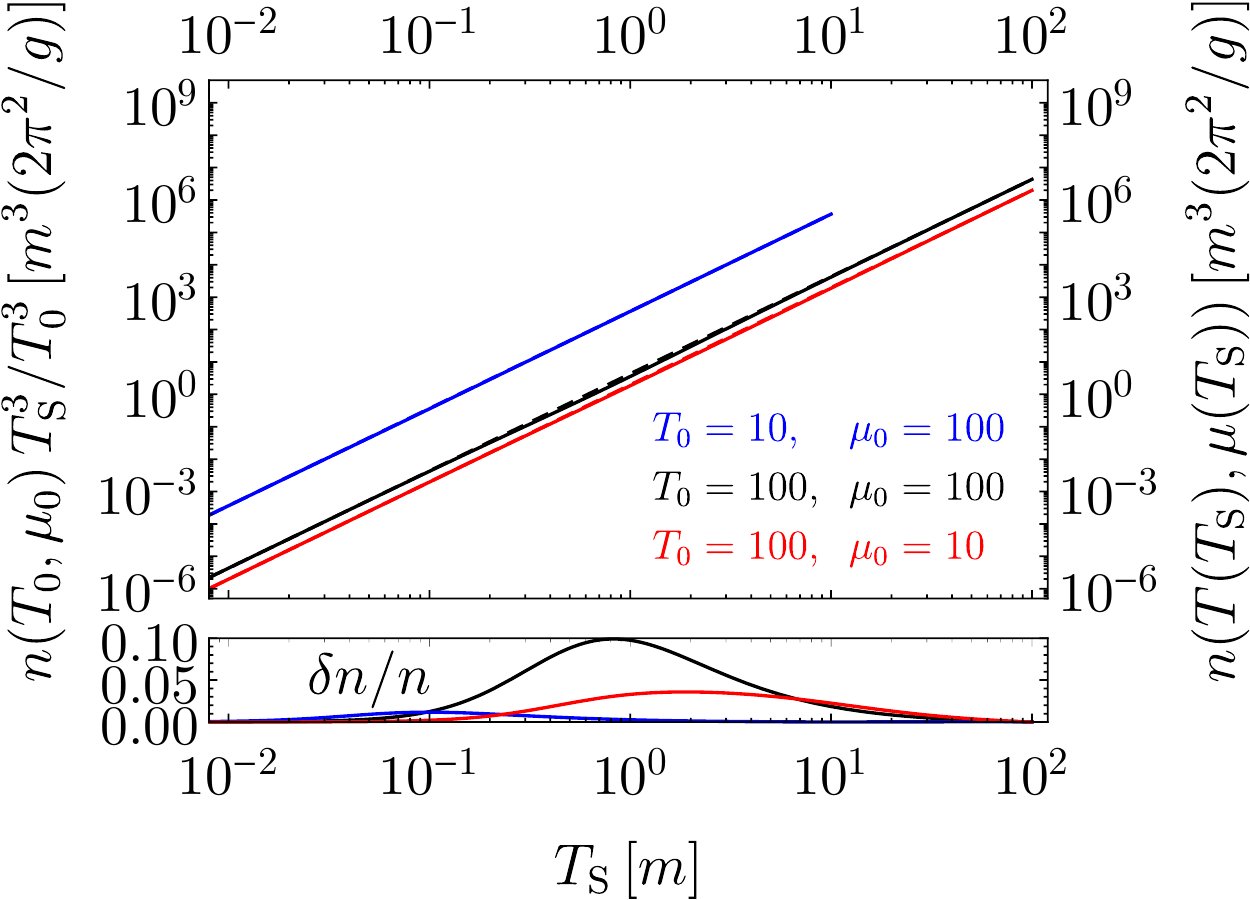}\label{fig:RhoDensity}
}
\caption{(a) Numerical values for $\rho$ as a solution to \eqref{eq:RhoCondition}. (b) Density obtained with $T(\TS)$ from \eqref{eq:TTsfitfunction} with the corresponding $\rho$, as well as $\mu(\TS)$ 
from \eqref{eq:muOfT} (solid) compared to the correctly scaled density (dashed) for different sets of initial values. 
The lower plot shows the relative error between the densities calculated with the aid of $T(\TS)$ and $\mu(\TS)$ and the correctly scaled densities.}
\end{figure}

Fixing the initial conditions, the only free parameter of both functions is $\rho$ in $T(\TS)$. 
We can fix $\rho$ by the requirement that the correct (scaled) density is obtained in the limit $\TS\rightarrow 0$.
In this limit, the density as a function of $T$ and $\mu$ can be approximated by \eqref{eq:ExactDensityDown}. Therefore, we plug $T(\TS)$ and $\mu(\TS)$ in \eqref{eq:ExactDensityDown}
and expand for small $\TS$. Requiring that the resulting expression coincides with the scaled density we find that $\rho$ is fixed by the transcendental relation
\begin{equation}\label{eq:RhoCondition}
 -\sqrt{\frac{\pi}{2}}\,\left(\frac{m\,\rho}{T_0}\right)^{3/2}\,\mathrm{Li}_{3/2}\left[-\exp\left(\frac{\mu_0^2-m^2}{2\,m\,T_0\,\rho}\right)\right]~=~\frac{n(T_0,\mu_0)}{T_0^3}\;.
\end{equation}
The initial density $n(T_0,\mu_0)$ can, for sufficiently large $T_0$ or $\mu_0$, be calculated via \eqref{eq:ExactDensityUp}.
Equation \eqref{eq:RhoCondition} has a unique solution in $\rho$. Numerical solutions for a range of initial parameters are 
shown in \Figref{fig:RhoValues}. As illustrated in \Figref{fig:RhoDensity}, we can indeed correctly reproduce the scaled density 
using the functions $\mu(\TS)$ and $T(\TS)$ for the effective chemical potential and the effective temperature, respectively. 
We find that the maximal discrepancy between the density computed with the effective functions and the correctly scaled initial density is $10\%$. 
The maximal error is reached in the region around $\TS\sim m$ only when $T_0\sim\mu_0$. In regions where $T_0$ is very different from $\mu_0$ the discrepancy is substantially reduced.

\section{Phase space density}

In the remainder of this work let us investigate the phase space density of massive and massless Fermi gases in an expanding volume. 
A measure for the dimensionless phase space density is $\varphi:=\langle \lambda \rangle n^{1/3}$, where $\langle \lambda\rangle$ denotes the expectation value of the de Broglie wavelength 
\begin{equation}
 \lambda~=~\frac{h}{p}~=~\frac{2\pi}{\sqrt{E^2-m^2}}\;.
\end{equation}
The phase space density or $\varphi$ compares the extension of a single particle to the volume per particle and, therefore, is a measure for the quantum degeneracy of a gas.
The thermal de Broglie wavelength is given by\footnote{%
It is straightforward to reproduce the commonly stated expressions for the thermal wavelength \mbox{$\lth\propto(mT)^{-1/2}$} and $\lth\propto T^{-1}$ 
in the non-- and ultra--relativistic limit of \eqref{eq:LambdaDefinition}.}
\begin{equation}\label{eq:LambdaDefinition}
 \lth~:=~\langle\lambda\rangle~=~-\frac{g}{2\pi^2}\frac{\,2\,\pi\,T^2}{\,n(T,\mu)}\left[\Li{2}{-z}-\frac{m}{T}\ln(1+z)\right]\;.
\end{equation}
Given $\lth$ in this form we can directly state $\varphi$ which is given by
\begin{equation}\label{eq:PSdensity}
 \varphi(m,T,\mu)~=~-\frac{g}{2\pi^2}\frac{\,2\,\pi\,T^2}{\,\left[n(T,\mu)\right]^{2/3}}\left[\Li{2}{-z}-\frac{m}{T}\ln(1+z)\right]\;.
\end{equation}
There is one very peculiar fact to note about $\varphi$: In contrast to the density $n(T,\mu)$, the two limits 
\begin{equation}\label{eq:NC}
 \lim_{T\rightarrow0}\varphi\quad\text{and}\quad\lim_{\mu\rightarrow m}\varphi
\end{equation}
do not commute when applied to $\varphi$. Physically this reflects the fact that while a gas in the limit $T\rightarrow 0$, $\mu>m$ will \textit{always} be maximally densely packed in phase space,
this maximal degeneracy will \textit{not} be achieved if the chemical potential is fixed at its lower bound $\mu=m$ while we turn off the temperature $T\rightarrow0$. 

Due to the non--commuting limitae \eqref{eq:NC} the phase space density at the point $(T,\mu)=(0,m)$ crucially depends on from which direction we approach this point.
If $T$ and $\mu$ can be varied independently of each other, as for the real temperature case, it is possible to fix one of them while the other is varied.
In this way, we can approach the point $(0,m)$ either from the ``$T$'' or from the ``$\mu$ direction'' resulting in the two different phase space densities.
To obtain the degenerate case we take $T\rightarrow 0$ (while $\mu>m$ is fixed) and the phase space density is given by
\begin{equation}\label{eq:maximalPSdensity}
 \varphi(m,T=0,\mu>m)~=~\left(\frac{g}{2\pi^2}\right)^{1/3}\,3^{2/3}\,\pi~=:~\varphi_{\mathrm{max}}~\;.
\end{equation}
This holds for massive and massless particles in the same way and constitutes a strict upper bound for $\varphi$ in all subsequent discussions.

\begin{figure}[t]
\subfigure[]{
\includegraphics[width=0.45\textwidth]{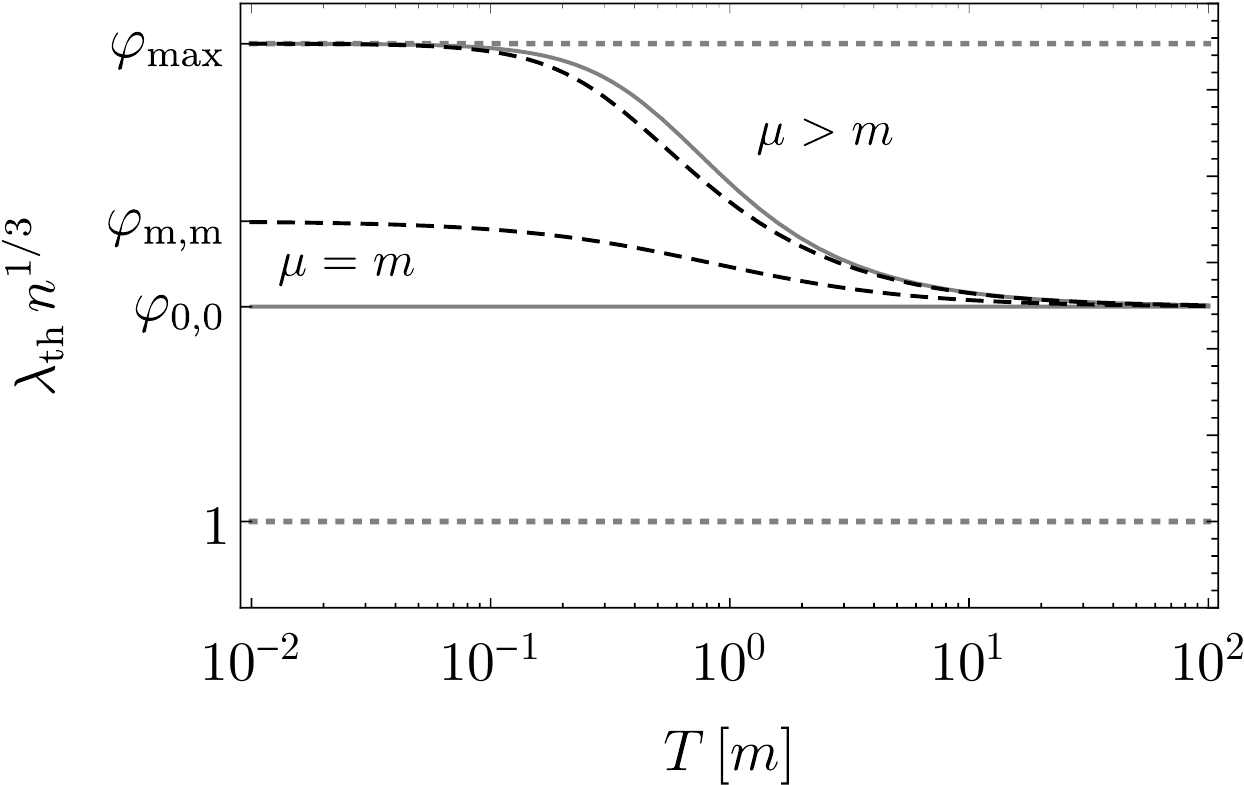}\label{fig:RealTPhaseSpace}
}
\subfigure[]{
\includegraphics[width=0.45\textwidth]{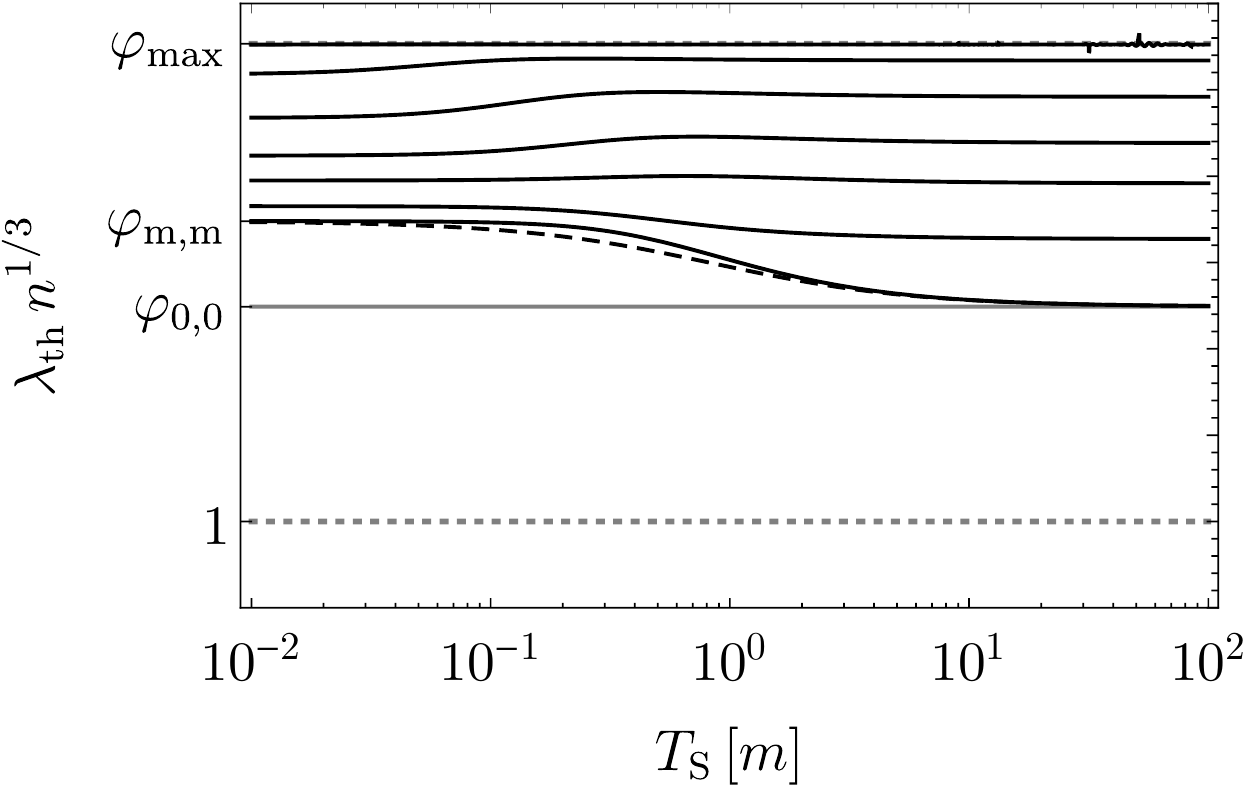}\label{fig:ScalingTPhaseSpace}
}
\caption{(a) Phase space degeneracy as function of the real temperature $T$ for the case of a massive (black, dashed) and massless gas (gray), for chemical potential $\mu=m$ (below) 
and for non--trivial chemical potential $\mu>m$ (above). (b) Phase space degeneracy as a function of the volume scaling $\TS$ for the massive case (black, solid) for different
initial parameters $\mu_0/T_0=\left\{ 0,1,2,3,5,10,50 \right\}$ (bottom to top). For comparison we also show the massless gas (gray) and the real temperature case (black, dotdashed).}
\end{figure}

For the opposite order of the limits it makes sense to discuss the massless case separately. 
Directly setting $m=0$ we use the density \eqref{eq:MasslessdensityMu} to compute
\begin{equation}\label{eq:masslessPSdensity}
 \varphi(0,T,\mu)~=~\left(\frac{g}{2\pi^2}\right)^{1/3}\,2^{1/3}\,\pi\frac{\,\,-\Li{2}{-\mathrm{e}^{\mu/T}}\hfill}{\left[-\Li{3}{-\mathrm{e}^{\mu/T}}\right]^{2/3}}\;,
\end{equation}
which holds for general $\mu$ and $T$. Using the limiting behavior of the polylogarithms (cf.\ Appendix \ref{app:Polylogs}) and requiring that $\mu>0$ we can take the limit $T\rightarrow 0$ 
to recover \eqref{eq:maximalPSdensity}, which is independent of $\mu$. On the other hand, taking the limit $\mu\rightarrow0$ first we find
\begin{equation}
 \varphi(m=0,T,\mu=0)~=~\left(\frac{g}{2\pi^2}\right)^{1/3}\,\frac{\pi^3}{3\cdot2^{1/3}\cdot(3\,\zeta(3))^{2/3}}~=:~\varphi_{0,0}~<~\varphi_{\mathrm{max}}\;.
\end{equation}
This implies that the phase space density for a massless gas with $\mu=0$ is constant and 
independent of the temperature. For a massive gas this conclusion does not hold. That is, letting $\mu=m$ the phase space density is in general a function of the temperature.
Nevertheless, we can use the density given in \eqref{eq:DensityDown} in order to compute the limit 
\begin{equation}
 \lim_{T\rightarrow 0}\varphi(m,T,\mu=m)~=~\left(\frac{g}{2\pi^2}\right)^{1/3}\,2\,\pi\,\ln(2)\,\xi~=:\varphi_{m,m}~<~\varphi_{\mathrm{max}}\;.
\end{equation}
These findings are summarized in \Figref{fig:RealTPhaseSpace}. 

Let us now consider the case of a decoupled gas in an expanding volume. 
In this case we are not dealing with $\varphi$ as a function of the real temperature and chemical potential,
but their effective counterparts $T(\TS)$ and $\mu(\TS)$. Letting $\TS\rightarrow0$ we realize that we have to deal with \textit{both} limits 
$T\rightarrow0$ and $\mu\rightarrow m$ simultaneously, i.e.\ we are approaching the point $(T,\mu)=(0,m)$ \textit{from a certain direction}.
This direction is fixed as a function of the initial conditions $(T_0,\mu_0)$. In the massless case, we are approaching $(0,m)$ on a straight line from $(T_0,\mu_0)$,
as is immediately clear from \eqref{eq:TMuScalingMassless}. 
In the massive case, however, we move on some curve parametrized by $\TS$, approximately given by $(T(\TS),\mu(\TS))$ with the functions \eqref{eq:muOfT} and \eqref{eq:TTsfitfunction} (cf.\ \Figref{fig:TMuScaling}).

The phase space density in dependence of $\TS$ is given by formally replacing $T\rightarrow T(\TS)$, $\mu\rightarrow\mu(\TS)$ in \eqref{eq:PSdensity}.
In the massless case eq.\ \eqref{eq:masslessPSdensity} holds and 
the temperature and chemical potential are both linearly rescaled. Therefore, the scaling dependence simply cancels and the 
phase space density is independent of $\TS$ and given by
\begin{equation}
 \varphi(0,T(\TS),\mu(\TS))~=~\left(\frac{g}{2\pi^2}\right)^{1/3}\,2^{1/3}\,\pi\frac{\,\,-\Li{2}{-\mathrm{e}^{\mu_0/T_0}}\hfill}{\left[-\Li{3}{-\mathrm{e}^{\mu_0/T_0}}\right]^{2/3}}\;.
\end{equation}
This justifies the statement that a massless gas behaves scale invariant in phase space, which is just another instance of the fact that the form of the spectrum is invariant under spatial expansion.
Depending on the initial conditions, $\varphi$ assumes a value in between $\varphi_{0,0}$ and
$\varphi_{\mathrm{max}}$.

In the massive case, $\varphi$ generally varies as a function of $\TS$. 
We can use the leading order behavior of $\mu(\TS)$ and $T(\TS)$ for low $\TS$ as well as the correctly scaled density in order to compute the
phase space density of a massive gas in the infinite volume limit
\begin{equation}
 \lim_{\TS\rightarrow 0}\varphi(m,T(\TS),\mu(\TS))~=~\left(\frac{g}{2\pi^2}\right)^{1/3}\,2\,\pi\,\rho\,\frac{m\,T_0}{n_0^{2/3}}\,\ln\left(1+\exp{\frac{\mu_0^2-m^2}{2\,T_0\,m\,\rho}}\right)\;.
\end{equation}
Depending on the initial conditions, $\varphi$ at infinite volume can reach all values in between $\varphi_{m,m}$ and $\varphi_\mathrm{max}$, \Figref{fig:ScalingTPhaseSpace}.

\section{Summary and Conclusion}

In this work we have discussed an ideal quantum gas of fermions in an expanding volume. 
At decoupling, the spectrum of the gas is characterized by an initial chemical potential and an initial temperature.
As the volume is rescaled, the spectrum of a decoupled gas generally deviates from an equilibrium Fermi--Dirac spectrum.
Nevertheless, for many cases the evolving spectrum can be characterized by an effective chemical potential $\mu(T_S)$ and an effective temperature $T(T_S)$, 
where the relative size of the volume, i.e.\ the ratio of scale factors, is expressed in terms of the scaling temperature $T_S\propto1/R$. 
The scaling temperature $\TS$ can be interpreted as the effective temperature of a massless background gas in the upscaled volume.

We have treated several different limiting cases. The special case of vanishing particle mass, 
the massive case with non--relativistic decoupling, as well as the massive case without self--interactions
are all classic textbook examples. We have added to this discussion the behavior of a completely degenerate massive gas,
with finite initial chemical potential and vanishing initial temperature, as well as the case of a massive 
and self--interacting gas which decouples relativistically and then is red--shifted until it becomes 
non--relativistic.

It has been shown that the assumption of an equilibrium spectrum after decoupling holds without limitations for the massless, 
and for the completely degenerate case. Furthermore, the assumption of an equilibrium spectrum holds to an excellent approximation
as long as the decoupled particles are \textit{either} all highly relativistic \textit{or} all non--relativistic.
However, for the particular case of a massive gas which decouples when it is relativistic and then is red--shifted until it 
becomes non--relativistic the form of the spectrum generally deviates from an equilibrium Fermi--Dirac spectrum.
Nevertheless, if there are sufficiently strong self--interactions within the decoupled gas
then the assumption of an equilibrium spectrum may be upheld even after the gas has become non--relativistic by 
red--shift \cite{Bernstein:1988bw,Lesgourgues:2013}.

For the completely degenerate case we have derived an exact form for the effective chemical potential $\mu(\TS)$, 
which scales down as the volume scales up and reaches the minimum value $\mu=m$ at infinitely upscaled volume.
This case could be realized in Nature as a cosmic background of non--thermal Dirac neutrinos \cite{Chen:2015dka}. 
The relic density of non--thermal cosmic background neutrinos is bounded above by $n_{\mathrm{nt}}\lesssim 217\cm^{-3}$ and their 
mean relic velocity has been calculated in this work to be
$\langle v_\text{C$\nu$B, nt} \rangle~\lesssim~572\left(0.1\,\ev/m_{\nu}\right)\text{km}\,\text{s}^{-1}$.

Secondly, we have treated the purely thermal case in which the chemical potential is fixed at the mass and the initial 
temperature of the massive gas coincides with the initial temperature of the massless background gas. 
We have used this case in order to extract a series expansion for the effective temperature $T(\TS)$. 
In addition, a phenomenological fit function has been found that interpolates $T(\TS)$ for all $\TS$. 
For the purely thermal case, the effective temperature after the gas has become non--relativistic is given by
$T_\mathrm{eff}\approx1.52\,\TS^2/m$ independently of the initial conditions. 
The mean relic velocity in this case is given by $\langle v(\TS)\rangle\approx2.12\,\TS/m$, and the spectrum
features a non--relativistic shape.
This should be contrasted to the strictly non--interacting (free--streaming, non--equilibrium) case in which the spectrum features a 
red--shifted relativistic shape and $\langle v(\TS)\rangle\approx3.15\,\TS/m$ at late times. 
We stress that the number density of particles as well as the mean energy per particle (to first order) coincides for both cases
irrespective of whether an equilibrium spectrum is preserved. 

We have also investigated the most general case with unconstrained initial chemical potential and temperature and have found that
it is well approximated by the obtained functions $\mu(\TS)$ and $T(\TS)$. 
The evolution with general initial condition is depicted in \Figref{fig:TMuScaling}.

Finally, we have computed the phase space densities for all cases comparing the real 
temperature case to the behavior of a gas in an expanding volume.

This study shows that in contrast to massless or degenerate particles,
the spectrum of decoupled massive particles is in general not invariant under the expansion of the universe.
The fact that $\langle \lambda \rangle n^{1/3}$ is much bigger than $1$ and even raises at late times shows that decoupled 
massive particles 
must be treated as a quantum gas throughout the evolution of the universe. In this sense, we hope that our study will also help to 
shed light on the question whether relic neutrinos condense and enter a superfluid phase at late times \cite{Ginzburg:1967a, Caldi:1999db, Kapusta:2004gi, Bhatt:2009wb, Azam:2010kw, Dvali:2016uhn}.
Furthermore, even though probably unobservable for generations to come, we remark that
an observation of the C$\nu$B spectrum will give insights into the non--standard self--interaction nature of neutrinos.
Therefore, this constitutes a further example of how it is possible to probe neutrino cosmology 
with terrestrial neutrino experiments (cf.\ e.g.\ \cite{Ghalsasi:2016pcj}). 
Our results could also be of importance for any kind of self--interacting or degenerate dark sector, 
examples being the cosmic neutrino background in the presence of non--standard interactions 
or certain cases of self--interacting Dark Matter.

\begin{acknowledgments}
I am grateful to G.\ Raffelt for enlightening discussions. Furthermore, I wish to thank M.\ Ratz, A.~Reinert, and A.\ Solaguren--Beascoa for comments on the manuscript,
N.\ Kaiser, D.\ Meindl, A.\ M\"utter, and W.\ Zwerger for helpful discussions, as well as V.\ Mukhanov for a useful correspondence.
This research was partially supported by the DFG cluster of excellence ``Origin and Structure of the Universe'' and by the 
German Science Foundation (DFG) within the SFB--Transregio TR33 ``The Dark Universe''.
\end{acknowledgments}

\appendix

\section{Asymptotic series for $\boldsymbol{n(T,\mu)}$}
\label{app:FermiDiracSeries}
In the integral for the density \eqref{eq:FermiSpectrum} we make the substitution $\epsilon=E-m$ and write the square root as a series expansion.
We can do this assuming either $\epsilon\ll m$ or $\epsilon\gg m$.
The resulting sums can formally be permuted with the integral resulting in
\begin{align}\label{eq:SeriesExpansion1}
 n(T,\mu)\,\frac{2\pi^2}{g}~=&~-T^3\,\sum_{n=0}^{\infty}\binom{1/2}{n}\left(\frac{2\,m}{T}\right)^n\left(\frac{3-n}{3-2n}\right)\Gamma(3-n)\,\Li{3-n}{-z}\,,\quad\text{and}\\
\label{eq:SeriesExpansion2}
 n(T,\mu)\,\frac{2\pi^2}{g}~=&~-\sqrt{2}\left(m\,T\right)^{3/2}\,\sum_{n=0}^{\infty}\binom{1/2}{n}\left(\frac{T}{2\,m}\right)^n\left(\frac{3+2n}{3-2n}\right)\Gamma(3/2+n)\,\Li{3/2+n}{-z}\;,
\end{align}
respectively. The first sum only gives a sensible approximation to the integral in case that $T\gg m$ and/or for the case that $\mu\gg m$.
Note, that in \eqref{eq:SeriesExpansion1} terms of the order $n>4$ can be divergent. 
The higher order terms could, in principle, be taken into account by adopting some regularization scheme, 
for example taking their Cauchy principal value. Nevertheless, this does not seem to improve the approximation obtained from the first four terms.
The second sum \eqref{eq:SeriesExpansion2} gives a good approximation to the integral in case that $T\ll m$ and $\mu\gtrsim m$. 
It does not suffer from the same problematic as \eqref{eq:SeriesExpansion1}.

Note that the existence of a closed form expression for the density \eqref{eq:FermiDiracIntegral} for all $T$ and $\mu$ would suggest a relation between polylogarithms of
degree $3-n$ and of degree $3/2+n$. We are not aware of any such relation.

\section{Series coefficients of $\boldsymbol{T(\TS)}$}
\label{app:TSupSeriesCoefficients}
The leading order expansion coefficients for $T(\TS)$ in \eqref{eq:TSExpansionUp} are given by
\begin{align}
 b^\mathrm{up}_1~=&~\frac{18\,n_0}{T_0}\,\frac{1}{27\,\zeta(3)\,T_0^2+2\,\pi^2\,m\,T_0+3\ln(2)\,m^2}\;,\\
 b^\mathrm{up}_2~=&~\frac{18\,n_0}{T_0}\,
		    \frac{-18\,\pi^2\,m\,n_0- 486\,\zeta(3)\,n_0\,T_0 + \left( 27\,\zeta(3)\,T_0^2 + 2\,\pi^2\,m\,T_0 + 3\ln(2)\,m^2  \right)^2}
		    {\left(27\,\zeta(3)\,T_0^2+2\,\pi^2\,m\,T_0 + 3\ln(2)\,m^2\right)^3}
\end{align}

\section{Limits for $\boldsymbol{\Li{s}{z}}$}
\label{app:Polylogs}
In order to derive and crosscheck some of the results of this work the following limits of polylogarithmic functions have proven to be very useful \cite{110}
\begin{align}\label{eq:polyloglimit}
 \lim_{\mathrm{Re}(x)\rightarrow\infty}~\Li{s}{-\mathrm{e}^x}~&=~-\frac{x^s}{\Gamma(s+1)}\;, \quad (s\neq -1,-2,-3,\dots)\;, \\
 \lim_{|z|\rightarrow0}~-\Li{s}{-z}~&=~z\;.
\end{align}

\bibliography{Orbifold}
\bibliographystyle{JHEP}
\end{document}